\documentclass[a4paper,11pt]{article}
\usepackage{jheppub} % for details on the use of the package, please see the JINST-author-manual
\usepackage{lineno}
\usepackage{comment}
\usepackage{tikz}
\usetikzlibrary{arrows}
\usepackage{xcolor}

\newcommand{\zm}[1]{{\color{blue} #1}}
\title{\boldmath Scalar-assisted magnetogenesis during the radiation-dominated epoch}

\author[b]{Alexander Ganz,}
\author[c]{Chunshan Lin,}
\author[a,c]{Mian Zhu}
\affiliation[a]{College of Physics, Sichuan University, Chengdu 610065, China}
\affiliation[b]{Institute for Theoretical Physics, Leibniz University Hannover, Appelstraße 2, 30167 Hannover, Germany}
\affiliation[c]{Faculty of Physics, Astronomy and Applied Computer Science, Jagiellonian University, 30-348 Krakow, Poland}
\emailAdd{zhumian@scu.edu.cn}
\emailAdd{mian.zhu@uj.edu.pl}

\abstract{We propose a novel mechanism to generate primordial magnetic fields (PMFs) strong enough to explain the observed cosmic magnetic fields. We employ a scalar field charged under U(1) gauge symmetry with a non-trivial VEV to provide an effective mass term to the EM field and thus break its conformal invariance. The primordial magneto-genesis takes place in the radiation dominated (RD) epoch, after the electroweak symmetry breaking (EWSB) phase. As a result, our mechanism is naturally free from the over-production of electric fields due to high conductivity in the RD epoch, and the baryon isocurvature problem which takes place only if magneto-genesis happens before the ESWB phase. In addition, we find that a significant amount of PMFs can be generated when the scalar field experiences a tachyonic phase. In this case, the scalar field is light and weakly coupled and has negligible energy density compared to the cold dark matter, hence the strong coupling problem and the back-reaction problem are also absent. Therefore, our model is free from the above-mentioned problems that frequently appear in other primordial magneto-genesis scenarios.}

\begin{document}
\maketitle
\flushbottom

\section{Introduction}
\label{sec:intro}

Observations have confirmed the ubiquitous nature of magnetic fields \cite{Durrer:2013pga}. Nonetheless, the origin of cosmic magnetic fields is still unclear. Observational signatures of magnetic fields in inter-galactic void regions \cite{Neronov:2010gir,Tavecchio:2010ja,Tavecchio:2010mk,Taylor:2011bn} imply a cosmological origin of large-scale magnetic fields, due to the difficulty to account for them in astrophysical mechanisms \cite{Bondarenko:2021fnn}. 

Understanding the origin of large-scale magnetic fields is a longstanding problem. The blazar observations suggest that the strength of magnetic fields with a coherence length of a few Mpc should be larger than $10^{-15}$ Gauss \cite{Finke:2015ona, MAGIC:2022piy}. It is generically believed that primordial magneto-genesis before the structure formation sources the seed magnetic field, which, after astrophysical dynamo and compression amplification mechanisms, forms the large-scale magnetic field observed today, see \cite{Brandenburg:2004jv,Subramanian:2015lua} for a review. The strength of the seed magnetic field remains unclear due to uncertainties in the details of the dynamo mechanisms, which typically vary from $10^{-12}$ G to $10^{-22}$ G \cite{Kandus:2010nw,Vachaspati:2020blt}. It is difficult to produce such magnetic seeds through the primordial magneto-genesis process within classical electromagnetism and conventional cosmology \cite{Turner:1987bw}.

Inflation is believed to be a major candidate for the production of a large-scale magnetic field. Over the past years, vast inflationary magneto-genesis models have been proposed in the literature \cite{Garretson:1992vt,Dolgov:1993vg,Gasperini:1995dh,Bamba:2003av,Bamba:2006ga,Martin:2007ue,Kanno:2009ei,Byrnes:2011aa,Barnaby:2012tk,Domenech:2015zzi,Patel:2019isj,Durrer:2022emo,Sasaki:2022rat}. All of these models share a key ingredient, namely, the breaking of conformal invariance  \cite{Turner:1987bw}. One of the most  well-studied ideas is the Ratra theory \cite{Ratra:1991bn}, in which a scalar field couples to the field strength $f(\phi)^2 F_{\mu \nu} F^{\mu \nu}$. Most of these models potentially suffer from the strong coupling problem \cite{BazrafshanMoghaddam:2017zgx} or the back reaction problem \cite{Green:2015fss}. Additionally, a recent study reveals an overproduction of baryon isocurvature perturbations in any magneto-genesis scenario above the electroweak (EW) scale \cite{Kamada:2020bmb}, arguably ruling out most inflationary magneto-genesis scenarios.

In this paper, we propose a novel mechanism free from all the above problems. 
We will instead consider a new scalar field charged under $U(1)$ gauge symmetry, and the conformal invariance is broken due to the effective mass term of the electromagnetic (EM) field generated due to a non-trivial vacuum expectation value (VEV) of the scalar. The magnetic field is then induced by the scalar currents in the early universe. In contrast to the previous work we assume that the scalar field is very light and is extremely weakly coupled, and therefore, the scalar particles are not thermalized before the EW scale. Very weakly coupled fermion particles (milli charged particles) have been extensively discussed in the literature as a potential candidate for dark matter \cite{GOLDBERG1986151,Cheung:2007ut} and could arise as a low-energy limit of a new light $U(1)^\prime$ gauge field which kinetically mixes with the Standard Model (SM) $U(1)$.

The baryon isocurvature problem becomes irrelevant as long as the magneto-genesis takes place below the EW scale. In the literature, the condition is conventionally fulfilled by working in a low-scale inflation scenario \cite{Ferreira:2014hma,Fujita:2016qab, Kobayashi:2019uqs,Yanagihara:2023qvx}, namely the reheating temperature is way below the EW scale, at the cost of being contrary to the standard thermal history of the universe. In our work, we instead assume that the magneto-genesis happens in the standard radiation dominated epoch \footnote{See also e.g., Ref. \cite{Papanikolaou:2023nkx}.}. The problem of baryon isocurvature can be evaded as long as the primordial magnetic field is generated after the electroweak symmetry breaking (EWSB) phase, without altering the standard thermal history of our universe.  Additionally, in the radiation dominated epoch, the universe is effectively a plasma, in which the induced electric field  is diluted away by the high conductivity. Thus, we will not need to worry about the over-production of electric fields. 

One may worry about the feasibility of producing sufficient magnetic fields in our setup, as the inflationary magnetogenesis in scalar QED theory is argued to be ineffective \cite{Mazzitelli:1995mp,Finelli:2000sh, Davis:2000zp, Bassett:2000aw,Emami:2009vd}. For instance, Ref. \cite{Emami:2009vd} reports a seed magnetic field of order $10^{-29}$ G on $\mathcal{O}(1) {\rm Mpc}^{-1}$ scale from the coupling of inflaton with gauge field, which is marginally acceptable assuming a highly efficient astrophysical amplification mechanism. In radiation dominated epoch things may become worse due to the existence of high conductivity. We will show that a significant amount of primordial magnetic fields can be generated in the radiation dominated epoch without the back-reaction problem, provided that the scalar field experiences a tachyonic growth phase. Thus, our formalism provides a viable mechanism for generating large scale magnetic field without all problems mentioned above.

We organize the paper as follows. We present our model in Sec. \ref{sec:model}. The general formalism of induced magnetic field is discussed in Sec. \ref{sec:formalism}. In Sec. \ref{sec:MFscalar}, we analyze the power spectrum of magnetic field when the scalar field is tachyonically amplified. We explicitly calculate the magnetic field when the magneto-genesis takes place before and after the electron-positron annihilation in Sec. \ref{sec:MFbefore} and Sec. \ref{sec:MFafter} respectively. We show that the seed magnetic field from the interaction with the scalar field can be compatible to observations and conclude in Sec. \ref{sec:conclusion}.

Throughout this manuscript, the scalar field $\phi$ takes mass dimension, and the fine structure constant is dimensionless such that the vector potential $A_{\mu}$ has mass dimension. The normal and conformal magnetic field shall carry $[M]^2$ dimension. The conformal time is denoted by $\tau$. Unless specified, a prime shall denote the differentiation with respect to $\tau$.

\section{Our Model}
\label{sec:model}
We work with the following action
\begin{equation}
\label{eq:action}
    S = \int d^4x \sqrt{-g} \left[ \frac{M_p^2}{2} \mathcal{R} - \frac{1}{2} \left( D_{\mu} \phi \right)^{\dagger} D^{\mu} \phi - V(\phi,\chi) + \mathcal{L}_{\chi} - \frac{1}{4}F_{\mu \nu}F^{\mu \nu} + \mathcal{L}_{bg} \right] ~.
\end{equation}
with $D_{\mu} \equiv \partial_{\mu} - i\varrho A_{\mu}$ and $\mathcal{R}/2$ the Einstein-Hilbert action. The parameter $\varrho$ is the effective coupling constant between the scalar field and the $U(1)$ gauge field. The term $\mathcal{L}_{\chi}$ represents the Lagrangian of an auxiliary scalar field $\chi$, which we shall explain in Sec. \ref{sec:effectiveV}. The complex scalar field $\phi$ is coupled to the auxiliary field $\chi$ via the interaction
\begin{equation}
\label{eq:potential}
    V(\phi,\chi) = \lambda^2 M_p \chi \vert\phi \vert^2 ~.
\end{equation}

The background geometry is described by the flat FLRW metric
\begin{equation}
    ds^2 = -dt^2 + a(t)^2 dx_idx^i = a(\tau)^2 (-d\tau^2 + dx_idx^i) ~,
\end{equation}
where $t$ is the cosmic time and $\tau \equiv \int dt/a$ the conformal time. As pointed out by \cite{Kamada:2020bmb}, magneto-genesis scenarios above the electroweak scale are strongly restricted by the baryon isocurvature problem. In light of this result, we will work in the radiation-dominated (RD) epoch throughout this paper. The scale factor and the Hubble parameter $H \equiv \dot{a}/a$ are parametrized as follows:
\begin{equation}
    a(\tau) = a_e \left( \frac{\tau}{\tau_e} \right) ~,~ H(\tau) = H_{\textrm{inf}} \frac{\tau_e^2}{\tau^2} ~,~ \tau > \tau_e > 0 ~,
\end{equation}
where $\tau = \tau_e$ represents the beginning of RD epoch and $a_e \equiv a(\tau_e)$ the scale factor at that time. We assume an instantaneous reheating process so the Hubble parameter at $\tau = \tau_e$ is identical to that in inflation epoch, which we denote as $H_{\rm inf}$. 

We will assume that both the scalar field $\phi$ and the auxiliary field $\chi$ are spectator fields whose energy density is negligible compared to other matter content in the early universe. The background radiation is thus governed by other matter content, which we abbreviated as $\mathcal{L}_{bg}$. We will examine the back-reaction from the scalar field $\phi$ and confirm this assumptions in Sec. \ref{sec:MFbefore} and Sec. \ref{sec:MFafter}.

\section{Formalism for the induced magnetic field}
\label{sec:formalism}
\subsection{Dynamical equations of motion}
\label{sec:MFeom}
The equations of motion (Eoms) for the vector potential $A_{\mu}$ is: 
\begin{equation}
\label{eq:EoMgeneral}
	\frac{1}{\sqrt{-g}}\partial_{\alpha} \left( \sqrt{-g} F_{\lambda \sigma} g^{\alpha \lambda} g^{\mu \sigma} \right) + \frac{i}{2} \varrho \left( \phi \nabla^{\mu} \phi^{\ast} - \phi^{\ast} \nabla^{\mu} \phi \right) - \varrho^2 A^{\mu} \phi^{\ast} \phi = 0 ~.
\end{equation}
In our scenario, energy budget stored in the scalar field is converted into the gauge field sector, and therefore, throughout the period of our interest, the last term in the above equation is negligible compared to the second term. On the other hand, as we will show in Sec. \ref{sec:MFafter}, the resolution of back-reaction problem requires a small coupling constant $\varrho$,  which can be another reason to neglect the $\varrho^2$ term. Adopting Weyl gauge $A_0 = 0$, the 0-th and $i$-th component of \eqref{eq:EoMgeneral} simplifies to
\begin{equation}
	\label{eq:0thEoM}
	\partial^i {A}_i^\prime = \frac{i\varrho}{2}a^2 \left( \phi {\phi}^{\ast \prime} - {\phi}^\prime \phi^{\ast} \right)~,
\end{equation}
\begin{equation}
\label{eq:ithEoM}
    A_i^{\prime \prime} - \partial_j^2 A_i + \partial_i (\partial^j A_j) = \frac{i}{2} \varrho a^2 \left( \phi \partial_i \phi^{\ast} - \phi^{\ast}  \partial_i \phi \right) ~.
\end{equation}
Notice that, a further choice of Coulumb gauge $\partial_i A^i = 0$ is incompatible with \eqref{eq:0thEoM} due to the presence of the source term. 

As we work in the RD epoch, the conductivity of our universe $\sigma$ becomes important, which can be approximated as \cite{Baym:1997gq}
\begin{equation}
\label{eq:sigma}
    \sigma \simeq  \frac{T}{\alpha^2 \ln (1/\alpha)} ~,
\end{equation}
where $T$ the temperature of the thermalized universe and $\alpha = 1/137$ the fine-structure constant. Eq. \eqref{eq:sigma} holds as long as Rutherford scattering dominates and determines the mean free path. Cosmological events such as $e^+e^-$ annihilation at $T \simeq 0.1 {\rm MeV}$, which we denote as $T_a \equiv  0.1 {\rm MeV}$, lead to a sudden drop of $\sigma$ \cite{Turner:1987bw}:
\begin{equation}
    \sigma = 10^{-10} \frac{m_e}{e^2} = 10^{-10} \frac{m_e}{4\pi \alpha} = 5.6 \times 10^{-13} {\rm GeV} ~.
\end{equation}
As a comparison, shortly before the annihilation event we have
\begin{equation}
    \sigma \simeq \frac{0.1 {\rm MeV}}{\alpha^2 \ln (1/\alpha)} = 3.8 \times 10^{-1} {\rm GeV} ~.
\end{equation}
Namely, the conductivity drops more than ten orders of magnitude.

The evolution of $A_{\mu}$ in the presence of a high conductivity is given by \cite{Turner:1987bw}:
\begin{equation}
    A_i^{\prime \prime} + 4\pi \sigma a A^{\prime}_i - \partial_j^2 A_i + \partial_i (\partial^j A_j) = \frac{i}{2} \varrho a^2 \left( \phi \partial_i \phi^{\ast} - \phi^{\ast}  \partial_i \phi \right) ~.
\end{equation}

\subsection{Electric and magnetic fields}
The electric and magnetic fields observed from a co-moving observer $u^{\nu} = (1,\vec{0})$ are
\begin{equation}
	E_{\mu} \equiv F_{\mu \nu} u^{\nu} = (0,-\dot{A}_i) ~,~ B_{\mu} \equiv \tilde{F}_{\mu \nu} u^{\nu} = \left( 0, \frac{1}{a} \epsilon_i^{jk} \partial_jA_k \right) ~.
\end{equation}
In terms of 3-d conformal EM fields, we have
\begin{equation}
	\mathcal{E}_i \equiv aE_i = - A_i^{\prime} ~,~ \mathcal{B}_i \equiv aB_i = \epsilon_i^{jk} \partial_jA_k ~.
\end{equation}

The EoM \eqref{eq:0thEoM} gives
\begin{equation}
	\partial^i \mathcal{E}_i = -\frac{i\varrho}{2}a^2 \left( \phi {\phi}^{\ast \prime} - {\phi}^\prime \phi^{\ast} \right) ~.
\end{equation}
and with the help of \eqref{eq:ithEoM} we have
\begin{equation}
	\mathcal{E}_i^{\prime} =  -A_i^{\prime \prime} = 4\pi \sigma a A_i^{\prime} - \partial_j^2 A_i + \partial_i (\partial^j A_j) - \frac{i}{2} \varrho a^2 \left( \phi \partial_i \phi^{\ast} - \phi^{\ast}  \partial_i \phi \right) ~,
\end{equation}
Finally, the dynamical equation for ${\cal E}_i$ and ${\cal B}_i$ are 
\begin{align}
   & \mathcal{E}_i^{\prime \prime} + 4\pi \sigma a \mathcal{E}_i^{\prime} - \partial^2 \mathcal{E}_i = S_{e,i} ~, \\
    \label{eq:Bconformaldynamic}
	&\mathcal{B}_i^{\prime \prime} + 4\pi \sigma a \mathcal{B}_i^{\prime} - \partial^2 \mathcal{B}_i = S_{b,i} ~,
\end{align}
where
\begin{align}
    S_{e,i} \equiv & i\varrho a^2 \left[ (\phi^{\ast \prime} + \mathcal{H} \phi^{\ast} ) \partial_i \phi - (\phi^{\prime} + \mathcal{H} \phi) \partial_i \phi^{\ast} \right] ~, \\
    S_{b,i} \equiv & \frac{i}{2} \varrho a^2 \epsilon_i^{jk} (\partial_j \phi \partial_k \phi^{\ast} - \partial_k \phi \partial_j \phi^{\ast}) = i\varrho a^2 \epsilon_i^{jk} \partial_j \phi \partial_k \phi^{\ast} ~.
\end{align}
Notably, the electric fields get diluted in a plasma, so we can simply focus on the magnetic field. We will come back to this issue in Sec. \ref{sec:electricfield}.

\if{}
\begin{align}
	\mathcal{E}_i^{\prime \prime} & \nonumber = - 4\pi \sigma a \mathcal{E}_i^{\prime} -\partial_j^2 A_i^{\prime} + \partial_i (\partial^j A_j^{\prime}) - \frac{i}{2}\varrho \left[ a^2 \left( \phi \partial_i \phi^{\ast} - \phi^{\ast}  \partial_i \phi \right) \right]^{\prime} \\
	& = - 4\pi \sigma a \mathcal{E}_i^{\prime} + \partial^2 \mathcal{E}_i + i\varrho a^2 \left( \phi^{\ast \prime} \partial_i \phi - \phi^{\prime} \partial_i \phi^{\ast} \right) - i \varrho a^2 \mathcal{H} \left( \phi \partial_i \phi^{\ast} - \phi^{\ast}  \partial_i \phi \right) ~. 
\end{align}
So, the dynamical equation for $\mathcal{E}_i$ is
\begin{equation}
	\mathcal{E}_i^{\prime \prime} + 4\pi \sigma a \mathcal{E}_i^{\prime} - \partial^2 \mathcal{E}_i = S_{e,i} ~, 
\end{equation}
\begin{equation}
	S_{e,i} = i\varrho a^2 \left[ (\phi^{\ast \prime} + \mathcal{H} \phi^{\ast} ) \partial_i \phi - (\phi^{\prime} + \mathcal{H} \phi) \partial_i \phi^{\ast} \right] ~.
\end{equation}
For the magnetic field,
\begin{equation}
	\mathcal{B}_i^{\prime} = \epsilon_i^{jk} \partial_jA_k^{\prime} = - \epsilon_i^{jk} \partial_j \mathcal{E}_k ~,
\end{equation}
\begin{align}
	\mathcal{B}_i^{\prime \prime} & \nonumber = - \epsilon_i^{jk} \partial_j \mathcal{E}_k^{\prime} = - \epsilon_i^{jk} \partial_j \left[ 4\pi \sigma a A_k^{\prime} - \partial^2 A_k + \partial_k (\partial^m A_m) - \frac{i}{2} \varrho a^2 \left( \phi \partial_k \phi^{\ast} - \phi^{\ast}  \partial_k \phi \right) \right] \\
	& = - 4\pi \sigma a \mathcal{B}_i^{\prime} + \partial^2 \mathcal{B}_i + \frac{i}{2} \varrho a^2 \epsilon_i^{jk}  \partial_j \left( \phi \partial_k \phi^{\ast} - \phi^{\ast}  \partial_k \phi \right) - \epsilon_i^{jk} \partial_j \partial_k (\partial^m A_m) ~.
\end{align}
The combination $\epsilon_i^{jk} \partial_j \partial_k$ vanishes, so we get
\begin{equation}
	\label{eq:Bconformaldynamic}
	\mathcal{B}_i^{\prime \prime} + 4\pi \sigma a \mathcal{B}_i^{\prime} - \partial^2 \mathcal{B}_i = S_{b,i} ~,
\end{equation}
\begin{equation}
    S_{b,i} \equiv \frac{i}{2} \varrho a^2 \epsilon_i^{jk} (\partial_j \phi \partial_k \phi^{\ast} - \partial_k \phi \partial_j \phi^{\ast}) = i\varrho a^2 \epsilon_i^{jk} \partial_j \phi \partial_k \phi^{\ast} ~.
\end{equation}
Notably, the electric fields get diluted in a plasma, so we can simply focus on the magnetic field. We will come back to this issue in Sec. \ref{sec:electricfield}.
\fi

\subsection{Dynamics of the magnetic field}
The magnetic field in the Fourier domain is
\begin{equation}
	\hat{\mathcal{B}}_i(\vec{x},\tau) = \int \frac{d^3 k}{(2\pi)^3} e^{i \vec{k} \cdot \vec{x}} \left[ \mathcal{B}_{\vec{k}}(\tau) \hat{b}_{\vec{k}} +  \mathcal{B}_{-\vec{k}}^{\ast}(\tau) \hat{b}_{-\vec{k}}^{\dagger} \right]\hat{e}_i(\hat{k}) ~.
\end{equation}
For simplicity we suppress the polarization state of magnetic field, and $\hat{e}_i$ now only indicates the ``direction'' of $\mathcal{B}$. The creation and annihilation operators satisfy
\begin{equation}
    [\hat{b}_{\vec{k}_1},\hat{b}_{\vec{k}_2}^{\dagger}] = (2\pi)^3 \delta^{(3)} (\vec{k}_1 - \vec{k}_2) ~.
\end{equation}

In the Fourier domain, the dynamical equation \eqref{eq:Bconformaldynamic} becomes
\begin{equation}
\label{eq:conformalBkeq}
    \mathcal{B}_k^{\prime \prime} + 4\pi \sigma a \mathcal{B}_k^{\prime} + k^2 \mathcal{B}_k = S_{b,\vec{k}}(\tau) ~,
\end{equation}
with the source term being
\begin{equation}
	\label{eq:source}
	S_{b,\vec{k}} \equiv \int d^3x S_{b,i} e^{-i\vec{k} \cdot \vec{x}} \equiv \int d^3x e^{-i\vec{k} \cdot \vec{x}} \left( i\varrho a^2 \epsilon_i^{jk} \partial_j \phi \partial_k \phi^{\ast} \right) ~.
\end{equation}

The scalar field $\phi$ can be decomposed into a homogeneous background $\bar{\phi}(t)$ and a perturbative part $\delta \phi(\vec{x},t)$:
\begin{equation}
    \phi(\vec{x},t) = \bar{\phi}(t) + \delta \phi(\vec{x},t) ~,
\end{equation}
with
\begin{equation}
    \delta \phi(\vec{x},t) = \int \frac{d^3 k}{(2\pi)^3} e^{i \vec{k} \cdot \vec{x}} \left( \phi_{\vec{k}} \hat{a}_{\vec{k}} + \phi^{\ast}_{-\vec{k}} \hat{a}^{\dagger}_{-\vec{k}} \right) ~,
\end{equation}
where the creation and annihilation operators satisfy
\begin{equation}
    [\hat{a}_{\vec{k}_1},\hat{a}_{\vec{k}_2}^{\dagger}] = (2\pi)^3 \delta^{(3)} (\vec{k}_1 - \vec{k}_2) ~.
\end{equation}
The source term in the operator form becomes
\begin{equation}
    \hat{S}_{b,\vec{k}} = i\varrho a^2 \int \frac{d^3p}{(2\pi)^3} (\vec{p} \times \vec{q}) \left( \phi_{\vec{p}} \hat{a}_{\vec{p}} + \phi^{\ast}_{-\vec{p}} \hat{a}^{\dagger}_{-\vec{p}} \right) \left( \phi_{\vec{q}} \hat{a}_{\vec{q}} + \phi^{\ast}_{-\vec{q}} \hat{a}^{\dagger}_{-\vec{q}} \right) ~,~ \vec{q} \equiv \vec{k} - \vec{p} ~,
\end{equation}
and the particular solution to the conformal magnetic field is
\begin{equation}
	\vec{\mathcal{B}}_k(\tau) = \int^{\tau} g_k (\tau;\tilde{\tau}) \vec{S}_{b,\vec{k}}(\tilde{\tau}) d\tilde{\tau} ~,
\end{equation}
where $g_k$ is the Green's function associated to \eqref{eq:conformalBkeq}. As we will show later, the fractional terms is much more important than the $k^2$ term, since the typical scale for inter-galaxy magnetic field is
\begin{equation}
\label{eq:kG}
    k_G/a_{\rm today} = 1 \rm{Mpc}^{-1} = 6.4 \times 10^{-39} GeV ~.
\end{equation}
The dynamical equation then simplifies to 
\begin{equation}
    \mathcal{B}_k^{\prime \prime} + 4\pi \sigma a \mathcal{B}_k^{\prime} = S_{b,\vec{k}}(\tau) ~.
\end{equation}

\subsection{Green's function in different epochs}
\subsubsection{Before annihilation}
Before the annihilation event, the electric conductivity is decided by \eqref{eq:sigma}. Since $T$ scales as $a^{-1}$ in the RD epoch, we define a new quantity
\begin{equation}
\label{eq:gamma}
    \gamma \equiv 2\pi \sigma a = \frac{2\pi}{\alpha^2 \ln (1/\alpha)} (aT)_{\text{RD}} ~,
\end{equation}
which is constant in the RD epoch. Specifically, the value of $aT$ can be evaluated at the radiation-matter equality epoch
\begin{equation}
\label{eq:aTRD}
    (aT)_{\text{RD}} = a_{\text{eq}}T_{\text{eq}} = \frac{a_{\text{today}}}{z_{eq} + 1} \times 1 \text{eV} \simeq 3.0 \times 10^{-13} \rm{GeV} ~,
\end{equation}
and accordingly
\begin{equation}
\label{eq:gammavalue}
    \gamma = \frac{2\pi}{\alpha^2 \ln (1/\alpha)} (aT)_{\text{RD}} = 7.3 \times 10^{-9} \rm{GeV} ~.
\end{equation}
It's easy to see that $\gamma \gg k_G$.  Therefore, the equation of motion for the magnetic field simplifies to 
\begin{equation}
    \mathcal{B}_k^{\prime \prime} + 2\gamma \mathcal{B}_k^{\prime} = S_{b,\vec{k}}(\tau),
\end{equation}
where $\gamma$ is a constant evaluated in the eq. (\ref{eq:gammavalue}), and the Green's function simplifies to
\begin{equation}
\label{eq:Greenbefore}
    g_k(\tau,\tau^{\prime}) \simeq \frac{\Theta(\tau - \tau^{\prime})}{2\gamma} \left[ 1 - e^{-2\gamma(\tau - \tau^{\prime})} \right] \simeq \frac{\Theta(\tau - \tau^{\prime})}{2\gamma} ~.
\end{equation}
\subsubsection{After annihilation}
After the annihilation event, we use
\begin{equation}
    4\pi \sigma a = 4\pi \sigma H_{\rm inf} a_e^2 \tau = 4\pi \sigma H_{\rm inf} \tau  \frac{(aT)_{\text{RD}}^2}{T_e^2} = \frac{4\pi^2 \sigma}{3M_p} \sqrt{\frac{g_{\ast}}{10}} (aT)_{\text{RD}}^2 \tau ~,
\end{equation}
where we've used the expression of energy density of background radiation
\begin{equation}
\label{eq:THrelation}
    \rho_{\rm bg} = 3H^2M_p^2 = \frac{g_{\ast} \pi^2 T^4}{30} ~,
\end{equation}
to write the $H_{\rm inf}$ as
\begin{equation}
    H_{\rm inf} = \frac{\pi \sqrt{g_{\ast}}T_e^2}{3\sqrt{10} M_p} ~.
\end{equation}
For convenience, we define the following constant
\begin{equation}
    \kappa \equiv \sqrt{\frac{2\pi^2 \sigma}{3M_p} \sqrt{\frac{g_{\ast}}{10}} (aT)_{\text{RD}}^2} = 3.0 \times 10^{-28} \times \left( \frac{g_{\ast}}{106.75} \right)^{\frac{1}{4}} {\rm GeV} ~,
\end{equation}
and one may verify that $\kappa \gg k_G/a_{\rm today}$ and the dynamical equation simplifies to
\begin{equation}
    \mathcal{B}_k^{\prime \prime} + 2 \kappa^2 \tau \mathcal{B}_k^{\prime} \simeq S_{b,\vec{k}}(\tau) ~,
\end{equation}
The corresponding Green's function become
\begin{equation}
    g_k(\tau,\tau^{\prime}) = \frac{\sqrt{\pi} e^{\kappa^2 \tau^{\prime 2}}}{2\kappa} \left[ {\rm Erf} (\kappa \tau) - {\rm Erf} (\kappa \tau^{\prime})\right] \Theta(\tau - \tau^{\prime}) ~.
\end{equation}
The conformal time is connected to the temperature $T$ through
\begin{equation}
\label{eq:tauTrelation}
    \tau = \frac{1}{aH} = \frac{T}{(aT)_{\text{RD}}} \frac{1}{H} =\frac{1}{(aT)_{\text{RD}}}\sqrt{\frac{10}{g_{\ast}}} \frac{3M_p}{\pi T} ~,~ d\tau = -\frac{1}{(aT)_{\text{RD}}}\sqrt{\frac{10}{g_{\ast}}} \frac{3M_p}{\pi T^2}dT ~,
\end{equation}

Thus, the conformal time after the annihilation event satisfies
\begin{equation}
    \tau  > \frac{1}{(aT)_{\text{RD}}}\sqrt{\frac{10}{g_{\ast}}}\frac{3M_p}{\pi  \times 0.1 {\rm MeV}} = \sqrt{\frac{106.75}{g_{\ast}}}\times 1.2 \times 10^{35} {\rm GeV}^{-1} ~,
\end{equation}
and we can see $\kappa \tau \gg 1$. The Green's function simplifies to
\begin{equation}
    g_k(\tau,\tau^{\prime}) \simeq \frac{1}{2\kappa^2 \tau^{\prime}} \left[ 1 - \frac{\tau^{\prime}}{\tau} e^{-\kappa^2 (\tau^2 - \tau^{\prime 2})} \right] \Theta(\tau - \tau^{\prime}) ~.
\end{equation}
Additionally, the exponential term quickly shrinks to zero as long as $\tau$ differs from $\tau^{\prime}$ even by a small ratio $\frac{\tau - \tau^{\prime}}{\tau^{\prime}} > 10^{-8}$. We're justified to ignore that exponential term as long as the tachyonic phase is long enough. In terms of temperature, the Green's function simplifies to 
\begin{equation}
    g_k(T,T^{\prime}) \simeq \frac{1}{2\kappa^2} \frac{(aT)_{\text{RD}} \pi T^{\prime} \sqrt{g_{\ast}}}{3\sqrt{10}M_p} \Theta(T^{\prime} - T) = \frac{4.7 \times 10^{23} T^{\prime}}{[{\rm GeV}]^2} \Theta(T^{\prime} - T) ~.
\end{equation}
Notice that $g_{\ast}$ within $\tau^{\prime}$ and $\kappa^2$ cancels with each other and the Green's function is independent of $g_{\ast}$.

\subsection{Power spectrum of magnetic field}

The two-point correlation function of magnetic field is
\begin{align}
	\langle \vec{\mathcal{B}}_{k_a}(\tau) \vec{\mathcal{B}}_{k_b}(\tau) \rangle & = \int^{\tau} \int^{\tau} d\tau_a d\tau_b g_{k_a} (\tau;\tau_a) g_{k_b} (\tau;\tau_b) \langle \vec{S}_{b,\vec{k}_a}(\tau_a) \vec{S}_{b,\vec{k}_b}(\tau_b) \rangle ~.
\end{align} 

We first evaluate the two-point correlation function of the source term:
\begin{align}
    & \quad \langle \vec{S}_{b,\vec{k}_1}(\tau_a) \vec{S}_{b,\vec{k}_2}(\tau_b) \rangle \nonumber = -\varrho^2 a^4 \int \frac{d^3p_{a}}{(2\pi)^3} \frac{d^3p_{b}}{(2\pi)^3} \left( \vec{p}_a \times  \vec{q}_a \right) \left( \vec{p}_b \times \vec{q}_b \right) \\
    & \times \phi_{\vec{p}_a} (\tau_a) \phi_{-\vec{q}_a}^{\ast} (\tau_a) \phi_{\vec{p}_b} (\tau_b) \phi_{-\vec{q}_b}^{\ast} (\tau_b) (2\pi)^6 \delta^{(3)} (\vec{p}_a + \vec{k}_2 - \vec{p}_b) \delta^{(3)} (\vec{k}_1 + \vec{k}_2 ) ~.
\end{align}
where we've used the fact
\begin{equation}
    \vec{p}_a + \vec{q}_a = \vec{k}_1 ~,~ \vec{p}_b + \vec{q}_b = \vec{k}_2 ~, 
\end{equation}
\begin{equation}
    \delta^{(3)} (\vec{p}_a + \vec{k}_2 - \vec{p}_b) \delta^{(3)} (\vec{p}_b + \vec{k}_1 - \vec{p}_a) = \delta^{(3)} (\vec{p}_a + \vec{k}_2 - \vec{p}_b) \delta^{(3)} (\vec{k}_1 + \vec{k}_2 ) ~.
\end{equation}

Following the conventional treatment of induced gravitational waves, we define the transfer function
\begin{equation}
    \phi_{\vec{k}} (\tau) \equiv \mathcal{T}_k(\tau;\tau_{\ast}) \phi_{\vec{k}} (\tau_{\ast}) ~,
\end{equation}
where $\tau_{\ast}$ is a reference time. The scalar power spectrum is defined as
\begin{equation}
    \langle \phi_{\vec{k}}(\tau) \phi_{\vec{k}^{\prime}} (\tau) \rangle = (2\pi)^3 \delta^{(3)} (\vec{k} + \vec{k}^{\prime}) \frac{2\pi^2}{k^3} P_{\phi}(k,\tau) ~.
\end{equation}
We also introduce the auxiliary variables
\begin{equation}
    k \equiv \vert \vec{k}_1 \vert ~;~ v_a \equiv p_a/k ~,~ v_b \equiv p_b/k ~,
\end{equation}
such that
\begin{align}
    & \quad \nonumber \langle \vec{S}_{b,\vec{k}_1}(\tau_a) \vec{S}_{b,\vec{k}_2} (\tau_b)\rangle = 4\pi^4 \varrho^2 a^4 (2\pi)^3 \delta^{(3)} (\vec{k}_1 + \vec{k}_2) \\
    & \times \int \frac{d^3p_a}{(2\pi)^3}  \frac{P_{\phi} (v_ak) P_{\phi} (v_bk)}{v_a^3v_b^3k^6} \vert \vec{k}_1 \times \vec{p}_a \vert^2 \mathcal{T}_{k_a}(\tau_a;\tau_e) \mathcal{T}_{k_b}(\tau_b;\tau_e) ~,
\end{align}

The power spectrum of conformal magnetic field at the time $\tau$ is defined as
\begin{equation}
	\langle \vec{\mathcal{B}}_{k_a}(\tau_a) \vec{\mathcal{B}}_{k_b}(\tau_b) \rangle \equiv \frac{2\pi^2}{k_a^3} (2\pi)^3 \delta^{(3)} (k_a + k_b) P_{\mathcal{B}} (k_a) ~,
\end{equation} 
so we have
\begin{align}
    P_{\mathcal{B}}(k) & \nonumber \equiv \frac{k^3}{2\pi^2}\int^{\tau} \int^{\tau} d\tau_a d\tau_b g_{k} (\tau;\tau_a) g_{k^{\prime}} (\tau;\tau_b) \langle \vec{S}_{b,\vec{k}}(\tau_a) \vec{S}_{b,\vec{k}^{\prime}}(\tau_b) \rangle \\
    & = \frac{1}{2} \varrho^2 k^4 \int dv_adv_b \frac{\vert \vec{v}_a \times \vec{v}_b \vert^2}{v_a^2v_b^2} P_{\phi} (v_ak) P_{\phi} (v_bk) \vert \mathcal{I} \vert^2 ~,
\end{align}
where $\mathcal{I}$ is the time integral
\begin{equation}
    \mathcal{I}(\tau,\tau_{\ast}) \equiv \int^{\tau} d\tau^{\prime} a^2(\tau^{\prime}) g_{k} (\tau;\tau^{\prime}) \mathcal{T}_{k_a}(\tau^{\prime};\tau_{\ast}) \mathcal{T}_{k_b}(\tau^{\prime};\tau_{\ast}) ~.
\end{equation}

The angular variable is determined by the implicit delta function, which appears when we integrate out $\vec{p}_b$:
\begin{equation}
    v_b^2 = 1 + v_a^2 - 2v_a \cos \theta ~,~ v_b \in (\vert 1-v_a \vert,1+v_a) ~.
\end{equation}
We further define auxiliary variables 
\begin{equation}
    s \equiv v_a + v_b ~,~ d \equiv \vert v_a - v_b \vert ~,
\end{equation}
and the power spectrum of normal magnetic field becomes
\begin{align}
    P_B(k,\tau) = \frac{P_{\mathcal{B}}(k,\tau)}{a^4} & \nonumber = 4 \varrho^2 \left( \frac{k}{a} \right)^4 \int_1^{\infty} ds \int_0^1 dd (1-d^2)(s^2-d^2)^{-2}(s^2-1) \\ 
    & \times P_{\phi} \left( \frac{s+d}{2} k, \tau_{\ast} \right) P_{\phi} \left( \frac{s-d}{2} k, \tau_{\ast} \right) \vert \mathcal{I}(\tau;\tau_{\ast}) \vert^2 ~,
\end{align}
where $\tau_{\ast}$ is a certain reference time. 

Sometimes, it would be convenient to write everything in terms of temperature. The time integral then becomes 
\begin{equation}
    \mathcal{I}(T;T_{\ast}) \equiv \int^{T} -\frac{3\sqrt{10}} {\pi} g_{\ast}^{-\frac{1}{2}} g_k(T;T^{\prime}) M_p (aT)_{\text{RD}} \frac{dT^{\prime}}{T^{\prime 4}}  \mathcal{T}_{k_a}(T^{\prime};T_{\ast}) \mathcal{T}_{k_b}(T^{\prime};T_{\ast}) ~,
\end{equation}
where we have used the following relation
\begin{equation}
   d\tau^{\prime}  a^2(\tau^{\prime}) = -\frac{1}{(aT)_{\text{RD}}}\sqrt{\frac{10}{g_{\ast}}} \frac{3M_p}{\pi T^{\prime 2}}dT^{\prime} \times \frac{(aT)_{\text{RD}}^2}{T^{\prime 2}} ~.
\end{equation}

\section{Amplification of scalar field through a tachyonic phase}
\label{sec:MFscalar}
\subsection{The necessity of amplification on the scalar field}
\label{sec:toy}
Let's start with a toy case where the $\phi$ field is simply a light scalar, i.e., $V(\phi) = \frac{1}{2} m^2 \vert\phi\vert^2$ with $m \ll H_{\rm inf}$. The scalar field acquires a nearly scale-invariant power spectrum in the inflation epoch
\begin{equation}
\label{eq:Pphiinf}
    P_{\rm \phi,inf} = \frac{H_{\rm{inf}}^2}{4\pi} ~.
\end{equation}
In the scalar QED theory, the scalar field $\phi$ couples to photons, which consequently induces a thermal mass for the scalar. The thermal mass can be estimated in the following. The magnitude of vector potential can be estimated by using
\begin{equation}
\label{eq:rhogamma}
    \rho_{\gamma} = \frac{1}{2} \left( \mathcal{E}^2 + \mathcal{B}^2 \right) = \frac{1}{2} \left[ (A_i^{\prime})^2 + (\epsilon_i^{jk} \partial_j A_k)^2 \right] \equiv \frac{g_{\gamma}}{2} \omega^2 (A_i)^2 ~,
\end{equation}
where $\rho_{\gamma}$ is the energy density of the background radiation. The structural constant $g_{\gamma}$ comes from our estimation $A_i^{\prime} \sim \omega A_i$ and $\partial_j A_k \sim kA_k = \omega A_k$. In principle, \eqref{eq:rhogamma} is valid only for photons with a specific frequency, namely:
\begin{equation}
    \frac{d\rho_{\gamma}}{dE} dE = \frac{g_{\gamma}(E)}{2} E^2 \frac{d(A_i(E))^2}{dE} dE ~.
\end{equation}
In vacuum state, one has $\partial_j A_k = kA_k = \omega A_k$. If further assumes the electric and magnetic fields contain similar energy of photon, then we see $g_{\gamma} = 2$. Although our case is much more complicated, $g_{\gamma} = 2$ would serve as a good estimation. Hereafter, we will take $g_{\gamma} = 2$ in the calculation.

For a gas of photons in equilibrium, the energy density per unit energy is
\begin{equation}
    \frac{d\rho_{\gamma}}{dE}dE = \frac{8\pi}{(2\pi)^3} \frac{E^3 dE}{e^{E/T} - 1}  ~,
\end{equation}
so that
\begin{equation}
    \rho_{\gamma} = \frac{T^4}{\pi^2} \int_0^{\infty} \frac{x^3dx}{e^x-1} = \frac{\pi^2}{15} T^4 ~.
\end{equation}

Similarly, assuming $g_{\gamma}$ varies slowly with respect to $E$, one has
\begin{equation}
    \frac{d(A_i(E))^2}{dE} dE = \frac{2}{g_{\gamma} E^2} \frac{d\rho_{\gamma}}{dE} dE = \frac{2}{g_{\gamma}} \frac{E dE}{e^{E/T} - 1} ~,
\end{equation}
which gives
\begin{equation}
    \langle A_{\mu} A^{\mu} \rangle = \langle A_i A^i \rangle \sim \frac{\pi^2}{3g_{\gamma}} T^2 ~.
\end{equation}
Therefore, during the radiation epoch, the scalar field acquires a thermal mass $m_T^2=\pi^2 \varrho^2 T^2/6$. The dynamical equation of the scalar field due to the thermal mass with $m_T^2\gg m^2$ is given by
\begin{equation}
    \phi_k^{\prime \prime} + \frac{2}{\tau} \phi_k^{\prime} + \left[k^2  + \frac{\pi^2}{6} \varrho^2 (aT)^2 \right] \phi_k = 0 ~.
\end{equation}
As we argued above, we can neglect the contribution of $k^2$ on the scale we concerned. For illustrative purpose, let's adopt one branch of general solution,
\begin{equation}
    \phi_k \propto \frac{\sin \left( \frac{\pi}{\sqrt{6}} \varrho (aT)_{\text{RD}} \tau \right)}{\tau} \propto T \sin \left( \sqrt{\frac{15}{g_{\ast}}} \frac{\pi \varrho M_p}{T} \right) ~.
\end{equation}
Notice that, the transfer function is insensitive to the relative coefficients of $\phi_k$. Take the beginning of RD epoch as the reference time, namely $\tau_{\ast} = \tau_e$, we write the transfer function as
\begin{align}
    {\cal T}(T,T_\ast) = \frac{T \sin \left( \sqrt{\frac{15}{g_{\ast}}} \frac{\pi \varrho M_p}{T} \right)}{T_e \sin \left( \sqrt{\frac{15}{g_{\ast}}} \frac{\pi \varrho M_p}{T_e} \right)} = \frac{\sqrt{g_{\ast}}\sin( \sqrt{\frac{15}{g_{\ast}} } \frac{\pi \varrho M_p}{T}) T }{\sqrt{15} \pi \varrho M_p} ~,
\end{align}
where we've used the fact $\sin x \simeq x$ in the $|x| \ll 1$ limit.
We see the scalar field is frozen as long as $m_T < a H$ and then starts to oscillate while decaying as $T$.
%We expect the scalar field decays as $\phi \propto T$ due to the $\varrho^2 A_{\mu} A^{\mu} \phi^2$ coupling. As a result the transfer function is $\mathcal{T}(T;T_{\ast}) = T/T_{\ast}$. 
Notably, the time integral and the power spectrum is scale-invariant in the scale we concerned, so the momentum integral becomes trivial:
\begin{equation}
\label{eq:momentumintres}
    \int_1^{\infty} ds \int_0^1 dd (1-d^2)(s^2-d^2)^{-2}(s^2-1) = \frac{1}{2} ~.
\end{equation}

We can easily calculate the magnetic field generated in our formalism. Before the annihilation event we have the Green's function \eqref{eq:Greenbefore}, so
\begin{align}
       \mathcal{I}(T;T_e) = & - \int^{T} \frac{3\sqrt{10}} {\pi \gamma \sqrt{g_{\ast}}} M_p (aT)_{\text{RD}} \frac{ d T^\prime}{T^{\prime 2}} \frac{g_{\ast} \sin \left( \sqrt{\frac{15}{g_{\ast}}} \frac{\pi \varrho M_p}{T^\prime}\right)^2 }{15 \pi^2 \varrho^2 M_p^2 } \nonumber \\
       = & - \frac{ \sqrt{5 g_{\ast}}}{ 20\sqrt{2} \pi^3 \gamma \varrho^2 M_p } (a T)_{\rm RD} \left( - \frac{2}{T} + \frac{\sin\left( 2 \sqrt{\frac{15}{2}} \frac{\pi \varrho M_p}{T}\right)}{\sqrt{\frac{15}{2}} \pi \varrho M_p} \right) \nonumber \\
       \simeq & 
       \frac{\sqrt{5 g_{\ast}} (a T)_{\rm RD} M_{\rm p}}{\sqrt{2} \pi \gamma T} \times \begin{cases}
           \frac{1 }{10  \pi^2 \varrho^2 M_p^2 } & \varrho M_p \gg T \\
           \frac{ 1 }{   T^2 } & \varrho M_p \ll T ~,
       \end{cases}
\end{align}
and 
\begin{align}
    P_B(T) & \nonumber = 2 \varrho^2 \left( \frac{k}{a} \right)^4 P_{\phi, \rm inf}^2 |\mathcal{I}(T;T_e)|^2 \\
    & \simeq  \left( \frac{k}{a} \right)^4 \frac{5g_{\ast} (a T)_{\rm RD}^2 H_{\rm inf}^4  }{16 \pi^4  \gamma^2 T^4} \times
    \begin{cases}
        \frac{T^2}{100 \pi^4 \varrho^2 M_p^2} & \varrho M_p \gg T \\
        \frac{\varrho^2 M_{\rm p}^2}{ T^2} 
 & \varrho M_p \ll T ~,
    \end{cases}
\end{align}
leading to a magnetic field strength at today 
\begin{align}
    B_{\rm today} \simeq 1.4 \times 10^{-72} g_{\ast}^{\frac{1}{2}} \left( \frac{k/a_{\rm today}}{1\,{\rm Mpc}^{-1}} \right)^2 \frac{H_{\rm inf}^2 }{T^2} \times
    \begin{cases}
        \frac{T}{10 \pi^2 \varrho M_{\rm p}} & \varrho M_p \gg T  \\
        \frac{\varrho M_{\rm p}}{T} & \varrho M_p \ll T 
    \end{cases}
    \, \left[{\rm G}\right] ~.
\end{align}
whose upper bound is 
\begin{align}
    & \quad B_{\rm today} \nonumber \leq 1.5 \times 10^{-71} \left( \frac{g_{\ast}}{106.75} \right)^{\frac{1}{2}} \left( \frac{k/a_{\rm today}}{1\,{\rm Mpc}^{-1}} \right)^2 \frac{H_{\rm inf}^2 }{T_a^2} \left[{\rm G}\right]\\
    & \leq 1.5 \times 10^{-39} \left( \frac{g_{\ast}}{106.75} \right)^{\frac{1}{2}} \left( \frac{k/a_{\rm today}}{1\,{\rm Mpc}^{-1}} \right)^2 \left( \frac{H_{\rm inf}}{10^{12} \left[{\rm GeV}\right]} \right)^2 \left( \frac{T_a}{10^{-4} \left[{\rm GeV}\right]} \right)^{-2} \left[{\rm G}\right] ~,
\end{align} 
which is not sufficient to explain the observations. 

After the annihilation event $T_a =0.1\,{\rm MeV}$, the conductivity drops. We set the reference time to be $T_{\ast} = T_a$ and get
\begin{align}
    \mathcal{I}(T;T_{a}) \equiv & \int_{T_a}^{T} - \frac{(aT)_{\text{RD}}^2 }{\kappa^2}   \frac{dT^{\prime}}{T^{\prime}} \frac{g_{\ast} \sin \left( \sqrt{\frac{15}{g_{\ast}}} \frac{\pi \varrho M_p}{T^\prime}\right)^2 }{30 \pi^2 \varrho^2 M_p^2 } \nonumber \\
    = & \frac{g_{\ast} (a T)_{\rm RD}^2}{60 \kappa^2 \pi^2 \varrho^2 M_p^2} \left( {\rm CosInt}(\sqrt{\frac{15}{g_{\ast}}} \frac{2 \pi \varrho M_p}{T_a}) -{\rm CosInt}(\sqrt{\frac{15}{g_{\ast}}} \frac{2 \pi \varrho M_p}{T}) + \log \frac{T_a}{T}  \right) \nonumber \\
    \simeq &  1.8 \times 10^{30} \frac{1}{\pi^2\varrho^2 M_p^2} 
    \begin{cases}
       \frac{15 \pi^2 \varrho^2 M_p^2}{g_{\ast} T^2} - \frac{15 \pi^2 \varrho^2 M_p^2}{g_{\ast} T_a^2}  & \varrho M_{p} \ll T < T_a \\
     \log \frac{T_a}{T} & \varrho M_p \gg T_a > T
    \end{cases}
    ~.
\end{align}
Accordingly
\begin{align}
    B_{\rm today} & \nonumber \simeq 10^{-30}  \left( \frac{k/a_{\rm today}}{1\,{\rm Mpc}^{-1}} \right)^2 \frac{H_{\rm inf}^2}{ \pi^2 \varrho M_p^2} \times
     \begin{cases}
     \frac{15 \pi^2 \varrho^2 M_p^2}{g_{\ast} T^2} - \frac{15 \pi^2 \varrho^2 M_p^2}{g_{\ast} T_a^2}  & \varrho M_{p} \ll T \\
     \log \frac{T_a}{T} & \varrho M_p \gg T_a
    \end{cases}
    \,\left[{\rm G}\right] \\
    &  \ll 10^{-30}  \left( \frac{k/a_{\rm today}}{1\,{\rm Mpc}^{-1}} \right)^2 \frac{H_{\rm inf}^2}{ T_a M_p} \times
     \begin{cases}
       \frac{15}{g_{\ast}} \frac{T_a}{T}   & \varrho M_{p} \ll T \\
     \log \frac{T_a}{T} & \varrho M_p \gg T_a
    \end{cases}
    \,\left[{\rm G}\right] ~.
\end{align}
Assuming $H_{\rm inf} = 10^{12} {\rm GeV}$, we have $H_{\rm inf}^2 /M_p T_a \sim 10^{9}$, which gives $B_{\rm today} \ll 10^{-21} \frac{T_a}{T} \left[{\rm G}\right]$. This is still insufficient to confront with observations unless we assign a ridiculously small $T$. We conclude that there must be an amplification for the scalar field to generate sufficient PMFs. 

\subsection{Setup and effective potential}
\label{sec:effectiveV}
As discussed before, we need to enhance the scalar field to generate sufficient PMFs. Let's introduce an auxiliary field $\chi$ which couples to the scalar $\phi$, and the potential for the scalar field $\phi$ including the thermal mass
\begin{equation}
\label{eq:Veffinteraction}
    V(\phi) = \lambda^2 M_p \chi \vert \phi \vert^2 + \frac{\pi^2}{12} \varrho^2 T^2 \vert \phi \vert^2 ~.
\end{equation}
We comment that the potential of $\phi$ shall include higher order terms, e.g., $\lambda_4^2 \phi^4$ for it to be bounded from below. Those term has to be suppressed by the factor $\phi/M_p$ at the magneto-genesis epoch for the validation of the effective potential \eqref{eq:Veffinteraction}. On the other hand, the interaction term $\lambda^2 M_p \chi \vert \phi \vert^2$ has negligible contribution to the dynamics of $\chi$ field since $\lambda$ is extremely small as we shall see later. Thus the dynamics of $\chi$ is approximately determined by $\mathcal{L}_{\chi}$. 

We shall adopt some ansatzs for the auxiliary field $\chi$. We assume that the $\chi$ field is initially in a local vacuum with $\langle \chi \rangle = -m^2/(\lambda^2  M_p) < 0$. Thus, when the universes cools down to a critical temperature
\begin{equation}
\label{eq:T1def}
    T_1 = \frac{2\sqrt{3}  m}{\pi \varrho} ~,
\end{equation}
the effective mass of $\phi$ field becomes negative and $\phi$ will experience a tachyonic growth. 
The growth stop when $\chi$ transitions to the positive expectation value $\langle \chi \rangle = M^2/(\lambda^2 M_p)$. The time scale is controlled by the form of the potential for $\chi$ and the respective coupling constants. 
%We assume that during the tachyonic growth the scalar field is enhanced up to a critical value $\phi \simeq v$. 
After the transition the scalar field $\phi$ will get a normal effective mass $M$ and the tachyonic behavior terminates. 
%\textcolor{red}{I modified the last paragraph. Please, check if you agree. I also started to renormalize the expectation value such that the tachyonic mass is given by $-m$ and the normal Mass by $M$. }

\subsection{Dynamics of scalar field}
Before the tachyonic phase, the scalar field $\phi$ locates at the true vacuum $\langle \phi \rangle = 0$ such that $\phi = \delta \phi$. The value of $\delta \phi$ can be estimated as $\langle \delta \phi \delta \phi \rangle = P_{\phi}$. When the temperature of universe is high, the effective potential of $\phi$ is simply $V(\phi) \simeq \pi^2 \varrho^2 T^2 \phi^2/12$ so that the thermal mass is given by $m_T^2 = \pi^2 \varrho^2 T^2/6$. As long as $m_T < H$ the scalar field remains frozen.  Using the relation \eqref{eq:THrelation}, we get
\begin{equation}
\label{eq:frozen}
    m_T < H ~\to~ \varrho < \sqrt{\frac{g_{\ast}}{15}} \frac{T}{M_p} ~,
\end{equation}
as the criteria for the scalar field to be frozen.

One may wonder whether the scalar field is still frozen after the annihilation epoch, where the temperature drops below 0.1 ${\rm MeV}$. We will evaluate the dynamics of scalar field by assuming that it's always frozen before the tachyonic phase for the following two reasons. First, as we show in Sec. \ref{sec:MFafteree}, a sizable magnetic field originates from a typically small value $\varrho \sim 10^{-25}$ and \eqref{eq:frozen} holds even when $T$ is of ${\rm keV}$ order. Second, from the computations in Sec. \ref{sec:MFbefore} and Sec. \ref{sec:MFafter} we confirm that the dynamics of scalar field has little impact on the magneto-genesis process, since any information before the tachyonic phase is diluted by the exponential growth phase.

After reaching the critical temperature, the dynamical equation of $\phi$ becomes
\begin{equation}
    \phi_k^{\prime \prime} + \frac{2}{\tau} \phi_k^{\prime} + \left[k^2  + \frac{\pi^2}{6} \varrho^2 (aT)^2 - 2 m^2 a^2 \right] \phi_k = 0 ~.
\end{equation}
We define two auxiliary constants
\begin{eqnarray}
    c_1^2 &\equiv& k^2  + \frac{\pi^2}{6} \varrho^2 (aT)^2 ~,\nonumber\\
    c_2 &\equiv&  m \frac{a_e}{\tau_e} = \frac{\lambda m}{H_{\rm inf} \tau_e^2} = \frac{\pi m}{3\sqrt{10}M_p} (a_{\rm eq} T_{\rm eq})^2 g_{\ast}^{-\frac{1}{2}} = 4.8  \times 10^{-46} \left( \frac{g_{\ast}}{106.75} \right)^{-\frac{1}{2}} m [{\rm GeV}] ~,
\end{eqnarray}
where $c_1$ is of dimension $[M]$ and $c_2$ has dimension $[M]^2$, and the dynamical equation simplifies to
\begin{equation}
    \phi_k^{\prime \prime} + \frac{2}{\tau} \phi_k^{\prime} + (c_1^2 - c_2^2 \tau^2) \phi_k = 0 ~,
\end{equation}
whose general solution is
\begin{equation}
\label{eq:phiksol}
    \phi_k = \mathcal{C}_1 \frac{e^{-\frac{c_2}{2} \tau^2}}{\tau} F_1 \left(\frac{1}{4} -\frac{c_1^2}{4c_2} ,\frac{1}{2},c_2\tau^2 \right) + \mathcal{C}_2 \frac{e^{-\frac{c_2}{2} \tau^2}}{\tau} \mathbf{H} \left( \frac{c_1^2}{2c_2} - \frac{1}{2},\sqrt{c_2}\tau \right) ~,
\end{equation}
where $F_1$ represents the Kummer confluent hypergeometric function and $\mathbf{H}$ represents the Hermitian polynomial. Notably, in the $\tau \to \infty$ limit, the branch of Hermitian polynomial approximates to a constant, and another branch has the following behavior
\begin{equation}
\label{eq:F1asymptotic}
    \frac{e^{-\frac{c_2}{2} \tau^2}}{\tau} F_1 \left(\frac{1}{4} -\frac{c_1^2}{4c_2} ,\frac{1}{2},c_2\tau^2 \right) \to \frac{\sqrt{\pi}}{\Gamma \left(\frac{1}{4} - \frac{c_1^2}{4c_2} \right)} \frac{e^{\frac{c_2}{2} \tau^2}}{\tau} (c_2 \tau^2)^{-\frac{1}{4} - \frac{c_1^2}{4c_2}} ~.
\end{equation}
Thus, the hypergeometric branch represents the exponential growing section since $\tau^2 \propto t$ and will be our focus. We calculate
\begin{equation}
\label{eq:c1c2condition}
    \frac{c_1^2}{c_2} = \frac{4.2 \times 10^{19} \varrho^2 + 1.2 \times 10^{-32} \left( \frac{k}{1{\rm Mpc}^{-1}} \right)^2}{ m/[{\rm GeV}]} \left( \frac{g_{\ast}}{106.75} \right)^{\frac{1}{2}} ~,
\end{equation}
which will be important in the following calculations.

\subsection{Matching condition and exponential amplification}

In this section, we aim to fix the integration constant ${\cal C}_1$ and ${\cal C}_2$ in Eq. \eqref{eq:phiksol}. First, the combination $c_2 \tau^2$ is a constant in inflationary epoch:
\begin{equation}
    c_2 \tau^2 = m \frac{a}{\tau} \tau^2 = ma\tau = \frac{m}{H} \ll 1 ~,
\end{equation}
Further, this quantity scales as $T^{-2}$ in the RD epoch, so
\begin{equation}
    c_2 \tau_1^2 = \frac{m}{H} \frac{T_{\rm inf}^2}{T_1^2} = \frac{3\sqrt{10} \pi}{\sqrt{g_{\ast}}} \frac{m}{T_1} \frac{M_p}{T_1} ~
\end{equation}
where $T_1$, defined by \eqref{eq:T1def}, is the critical temperature the tachyonic phase begins and we denote the corresponding conformal time as $\tau_1$. Later in Sec. \ref{sec:MFafteree}, we will show that $m M_p \ll T_1^2$, enabling us to set $c_2 \tau_1^2 \ll 1$ and write \eqref{eq:phiksol} as
\begin{equation}
    \phi_k(\tau) = \frac{\mathcal{C}_1}{\tau} + \frac{\mathcal{C}_2 \sqrt{\pi}}{2\sqrt{2} } \left( \frac{2}{\tau \Gamma\left( \frac{3}{4} \right)} - \frac{\sqrt{c_2}}{\Gamma\left( \frac{5}{4} \right)}  \right) + \mathcal{O}(\tau) ~,~ \tau \sim \tau_1 ~.
\end{equation}

As the scalar field is frozen till the critical epoch the initial conditions are simply given by
\begin{align}
\label{eq:initialcondition}
    \phi_k = \frac{H_{\rm inf}}{2 \sqrt{\pi}}, \qquad \phi_k^\prime =0
\end{align}
A direct match of the value of $\phi_k$ and $\phi_k^{\prime}$ at the critical time $\tau_1$ indicates that the $\tau^{-1}$ term must vanish, so
\begin{equation}
    \mathcal{C}_1 = - \frac{\mathcal{C}_2 \sqrt{\pi}}{2\sqrt{2} }  \frac{2}{\Gamma\left( \frac{3}{4} \right)} ~,~  - \frac{\sqrt{c_2}}{\Gamma\left( \frac{5}{4} \right)} \frac{\mathcal{C}_2 \sqrt{\pi}} {2\sqrt{2} } = \frac{H_{\rm inf}}{2 \sqrt{\pi}} ~,
\end{equation}
which tells
\begin{align}
    {\cal C}_1 \simeq  \frac{\Gamma\left( \frac{5}{4} \right)}{\Gamma\left( \frac{3}{4} \right)} \frac{H_{\rm inf}}{\sqrt{\pi c_2}} ~. 
\end{align}
We can understand the fact from an alternative perspective. At the beginning of RD epoch, the scalar fluctuations we concern are deeply outside the horizon and are approximately frozen. By frozen, we mean $\phi_k^{\prime}$ should be much less than $\mathcal{H}^2$. The term proportional to $\tau^{-1}$ in $\phi_k$ would lead to a $\tau^{-2}$ term in $\phi_k^{\prime}$, which is comparable to $\mathcal{H}^2 \propto \tau^{-2}$. Therefore, the frozen of scalar fluctuation leads to the vanish of coefficients of the $\tau^{-1}$ term.

When $c_2 \tau^2 > 1$, the exponential growth begins and we can use the asymptotic expression of Hypergeometric function to write
\begin{equation}
    \phi_k \simeq \frac{\Gamma\left( \frac{5}{4} \right)}{\Gamma\left( \frac{3}{4} \right)}  \frac{H_{\rm inf}}{\Gamma \left(\frac{1}{4} - \frac{c_1^2}{4c_2} \right)} e^{\frac{c_2}{2} \tau^2} (c_2 \tau^2)^{-\frac{3}{4} - \frac{c_1^2}{4c_2}} ~,
\end{equation}

Define an auxiliary parameter 
\begin{equation}
\label{eq:c3}
    c_3^2 \equiv \frac{3\sqrt{10} \pi}{2\sqrt{g_{\ast}}} M_p m =  1.7 \times 10^{19} \left( \frac{g_{\ast}}{106.75} \right)^{-\frac{1}{2}} m  [{\rm GeV}] ~,
\end{equation}
so that $c_2\tau^2/2 = c_3^2/T^2$ and we can write
\begin{align}
    \label{eq:phikasymp}
    \phi_k \simeq 0.20 \times 2^{-\frac{3}{4} - \frac{c_1^2}{4c_2}} H_{\rm inf} \frac{\Gamma \left(\frac{1}{4} \right)}{\Gamma \left(\frac{1}{4} - \frac{c_1^2}{4c_2} \right)} \left( \frac{T}{c_3} \right)^{\frac{3}{2} + \frac{c_1^2}{2c_2}}  e^{\frac{c_3^2}{T^2}} ~,
\end{align}
up to an overall phase factor, and in terms of $T$, one can easily check that $\phi_k$ has the dimension of mass. The validity of \eqref{eq:phikasymp} starts to break down when $\chi$ deviates from the local vacuum. Since initially $\langle \phi \rangle = 0$, the new VEV of scalar field is decided by the fluctuations with the longest wavelength, namely $\langle \phi \rangle = \phi_k(k = 0)$. As we shall reveal in the following sections, $c_1^2 \ll c_2$ in the limit $k \to 0$, thus the reference slice (labeled by $T = T_2$) when the tachyonic phase terminates is indicated by
\begin{equation}
\label{eq:vlambdam}
    v =  \frac{0.38}{\pi} H_{\rm inf} \left( \frac{T_2}{c_3} \right)^{3/2}  e^{\frac{c_3^2}{T_2^2}} ~.
\end{equation}

\subsection{Scale dependence of scalar power spectrum}
\label{sec:scaledependence}
Finally, let us discuss the scale dependence of scalar power spectrum, which is crucial to the evaluation of momentum integral. The k-dependence of power spectrum originates from $c_1(k)$, so we first write down the $c_1$ dependence of $P_{\phi}$
\begin{equation}
    P_{\phi}(c_1) \propto \frac{1}{\Gamma^2 \left(\frac{1}{4} - \frac{c_1^2}{4c_2} \right)} \left( \frac{T^2}{2c_3^2} \right)^{\frac{c_1^2}{2c_2}} ~.
\end{equation}
It is straightforward to see that $P_{\phi}$ monotonously decreases with respect to the ratio $c_1^2/c_2$ and $P_{\phi} = 0$ when $c_1^2 = c_2$. This fact can be understood as follows. The exponential growth begins when $c_2 \tau^2 > 1$. However, if in this case $c_1^2 > c_2$, then $m_{\rm eff}^2 = c_1^2 - c_2^2 \tau^2 > 0$. This means scalar fluctuations with $c_1^2 > c_2$ cannot feel the exponential growth, and the associated $P_{\phi}$ is negligible. We conclude that $P_{\phi}(k) \simeq 0$ when the wavenumber satisfies $c_1^2(k) \geq c_2$.

Thus, to determine the scale dependence of $P_{\phi}$, it's important to emphasis the $c_1(k)$ relation. Utilizing \eqref{eq:c1c2condition}, we find that \zm{$c_1^2/c_2$ is approximately a constant when $k \ll 5.9 \times 10^{25} \varrho$ ${\rm Mpc}^{-1}$ and $c_1^2/c_2$ scales as $k^2$ when $k \gg 5.9 \times 10^{25} \varrho $ ${\rm Mpc}^{-1}$.} 

\begin{figure}[ht]
    \centering
    \includegraphics[width=0.8\linewidth]{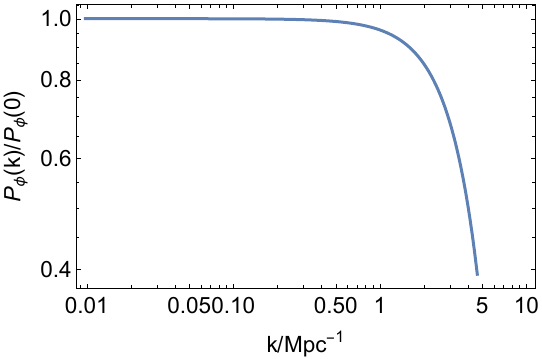}
    \caption{The scalar power spectrum as a function of $k$. We take $\rho = 1.7 \times 10^{-26}$ and $m = 1.2 \times 10^{-30}$ [{\rm GeV}] for illustrative purpose such that the critical scale is $k_c = 5.9 \times 10^{25} \varrho {\rm Mpc}^{-1} \simeq 1 {\rm Mpc}^{-1}$.}
    \label{fig:Pphik}
\end{figure}
With all above considerations, we see that $P_{\phi}$ is approximately a constant before a critical scales, and decreases rapidly, and illustrate the scalar power spectrum in Fig. \ref{fig:Pphik} to verify our conclusion. It's easy to see that the critical scale is determined by $k_c = 5.9 \times 10^{25} \varrho {\rm Mpc}^{-1}$, dependent solely on model parameter $\varrho$. Large scales with $k < k_c$ are approximately scale-invariant with $P_{\phi}(k) \simeq v^2$, while the scalar power spectrum quickly diminishes on small scales with $k > k_c$. For simplicity, we may take $P_{\phi}(k) \simeq 0$ when $k > k_c$, and write
\begin{equation}
\label{eq:Pphistep}
    P_{\phi} = v^2 \Theta(k_c - k) ~,~
\end{equation}
where $\Theta$ is the Heaviside function. 

\section{Magneto-genesis before the annihilation}
\label{sec:MFbefore}
Now we have successfully amplified the scalar field in the RD epoch, the next step is to calculate the induced magnetic field. In this section, we first work out the case when magneto-genesis happens before the $e^+e^-$ annihilation epoch. In this section, we compute the magnetic field generated during the tachyonic phase before the annihilation event. As the evaluations of magnetic field in different cases follow a similar logic, We will first illustrate the technique details extensively in Sec. \ref{sec:duringtacbefore}, and include only main results in the subsequent cases.

\subsection{Magnetic field generated during the tachyonic phase}
\label{sec:duringtacbefore}
\subsubsection{Analytical expression of magnetic power spectrum}
We set the reference slice of time integral to be $T = T_2$, and the power spectrum of induced magnetic field at $T = T_2$ is
\begin{align}
    P_B(k,T) & \nonumber = 4 \varrho^2 \left( \frac{k}{a} \right)^4 \int_1^{\infty} ds \int_0^1 dd (1-d^2)(s^2-d^2)^{-2}(s^2-1) \\ 
    & \times P_{\phi} \left( \frac{s+d}{2} k, T_2 \right) P_{\phi} \left( \frac{s-d}{2} k, T_2 \right) \vert \mathcal{I}(T_2,T_1;T_2) \vert^2 ~,
\end{align}
\begin{equation}
    \mathcal{I}(T_2,T_1;T_2) \equiv \int^{T_2}_{T_1} -\frac{3\sqrt{10}} {2\pi \gamma \sqrt{g_{\ast}}} M_p (aT)_{\text{RD}} \frac{dT}{T^4}  \mathcal{T}(T;T_2) \mathcal{T}(T;T_2) ~.
\end{equation}

There is a tricky point. The argument of Hypergeometric function  ranges from $c_2 \tau^2 \ll 1$ to $c_2 \tau^2 \to \infty$, where the function exhibit different asymptotic behavior. It would be difficult to perform the time integral without any simplification of special functions. Nonetheless, before the exponential amplification, the generated magnetic field is too small to be compatible with observations. Thus, we can safely calculate the transfer function using the asymptotic form of the Hypergeometric function at $c_2 \tau^2 \to \infty$. It's easy to see the transfer function during this period simplifies to
\begin{equation}
\label{eq:transfermg}
    \mathcal{T}(T;T_2) = \frac{\phi_k(T)}{\phi_k(T_2)} \simeq \left( \frac{T}{T_2} \right)^{\frac{3}{2} + \frac{c_1^2}{2c_2}} e^{c_3^2 \left( T^{-2} - T_2^{-2} \right)} ~.
\end{equation}
The time integral becomes
\begin{align}
\label{eq:timeintbefore}
    \mathcal{I} & \nonumber = \int^{T_2}_{T_1} -\frac{3\sqrt{10}} {2\pi \gamma \sqrt{g_{\ast}}} M_p (aT)_{\text{RD}} \frac{dT}{T_2^4} \left( \frac{T}{T_2} \right)^{\frac{c_1^2}{c_2} - 1} e^{2c_3^2 \left( T^{-2} - T_2^{-2} \right)} \\
    & \nonumber = \frac{3\sqrt{10}} {4\pi \gamma \sqrt{g_{\ast}}} M_p (aT)_{\text{RD}} \left[ \left( \frac{-c_3^2}{2T^2} \right)^{\frac{c_1^2}{2c_2}} T e^{- \frac{2c_3^2}{T_2^2} }  \left( \frac{T}{T_2} \right)^{\frac{c_1^2}{c_2} - 1} \Gamma \left( \frac{c_1^2}{2c_2} , - \frac{2c_3^2}{T^2} \right) \right]_{|_{T_1}^{T_2}} \\
    & \nonumber \simeq \frac{3\sqrt{10}} {4\pi \gamma \sqrt{g_{\ast}}} \frac{M_p}{T_2^4} (aT)_{\text{RD}} \left[ \frac{T_2^3}{2c_3^2} - \frac{T_1^3}{2c_3^2} \left( \frac{T_1}{T_2} \right)^{\frac{c_1^2}{c_2} - 1} e^{2c_3^2 \left( T_1^{-2} - T_2^{-2} \right)} \right] \\
    & \nonumber = \frac{3\sqrt{10} \frac{M_p}{T_2^4} (aT)_{\text{RD}}} {8\pi \gamma T_2 c_3^2 \sqrt{g_{\ast}} } \times \left[ 1 - \left( \frac{T_1}{T_2} \right)^{\frac{c_1^2}{c_2} + 2} e^{2c_3^2 \left( T_1^{-2} - T_2^{-2} \right)} \right] \\
    & \simeq \frac{5.0 \times 10^{-6}}{ m T_2} \times \left[ 1 - \left( \frac{T_1}{T_2} \right)^{\frac{c_1^2}{c_2} + 2} e^{-2c_3^2/T_2^2} \right] ~,
\end{align}
where we used the fact $c_3 \gg T_1$ to admit the exponential amplification and also $T_1 \gg T_2$ .

From the argument in Sec. \ref{sec:scaledependence}, the power spectrum $P_{\phi}$ can be non-trivial only if $c_1^2 \ll c_2$. Thus, we are justified to collect the contributions of momentum integral when this limit is satisfied, where \eqref{eq:timeintbefore} simplifies to
\begin{equation}
    \mathcal{I} \simeq \frac{5.0 \times 10^{-6}}{ m T_2} \times \left[ 1 - \left( \frac{T_1}{T_2} \right)^{2} e^{-2c_3^2/T_2^2} \right] ~.
\end{equation}
In addition, the expression of transfer function \eqref{eq:transfermg} is valid only if $T_1 < c_3$, otherwise \eqref{eq:phikasymp} is not applicable anymore. Thus
\begin{equation}
    0 < \left( \frac{T_1}{T_2} \right)^{2} e^{-2c_3^2/T_2^2} \leq \left( \frac{c_3}{T_2} \right)^{2} e^{-2c_3^2/T_2^2} \ll 1 ~,
\end{equation}
and we finally arrive at  
\begin{equation}
    \mathcal{I} \simeq \frac{5.0 \times 10^{-6}}{ m T_2} ~,
\end{equation}
which greatly simplifies the power spectrum to
\begin{align}
    P_B(k,T) \nonumber \simeq \frac{10^{-10}}{m^2T_2^2} \varrho^2 \left( \frac{k}{a} \right)^4 & \int_1^{\infty} ds \int_0^1 dd (1-d^2)(s^2-d^2)^{-2}(s^2-1) \\ 
    & \times P_{\phi} \left( \frac{s+d}{2} k, T_2 \right) P_{\phi} \left( \frac{s-d}{2} k, T_2 \right) ~.
\end{align}

To proceed, we need to use the expression of $P_{\phi}$, Eq. \ref{eq:Pphistep}. Apparently, the momentum integral vanished when $k/k_c \geq 2$. In the case $k/k_c < 2$, the step function in \eqref{eq:Pphistep} gives
\begin{equation}
    0 \leq \frac{k_1}{k_c} \leq 1 ~,~  0 \leq \frac{k_2}{k_c} \leq 1 \to 0 \leq s \pm d \leq \frac{2k_c}{k} ~,
\end{equation}
along with the constraints $s \geq 1$, we can translate the step function by a modified integration range of momentum integral. Let us define 
\begin{equation}
    \mathcal{I}_M \equiv \int_1^{\infty} ds \int_0^1 dd (1-d^2)(s^2-d^2)^{-2}(s^2-1) \Theta(k_c - k_1) \Theta(k_c - k_2) ~,
\end{equation}
when $1 \leq k/k_c \leq 2$, we have
\begin{align}
    \mathcal{I}_M = \int_1^{\frac{2k_c}{k}} ds \int_0^{\frac{2k_c}{k}-s} dd (1-d^2)(s^2-d^2)^{-2}(s^2-1) ~,
\end{align}
and when $0 \leq k/k_c \leq 1$, we have
\begin{align}
    \mathcal{I}_M & \nonumber = \left( \int_1^{\frac{k_c}{k}} ds \int_0^{{\rm min}(s,1)} dd +  \int_{\frac{k_c}{k}}^{\frac{2k_c}{k}} ds \int_0^{{\rm min} (1,\frac{2k_c}{k} - s)} dd \right) (1-d^2)(s^2-d^2)^{-2}(s^2-1) \\
    & = \left( \int_1^{\frac{2k_c}{k} - 1} ds \int_0^{1} dd +  \int_{\frac{2k_c}{k} - 1}^{\frac{2k_c}{k}} ds \int_0^{ \frac{2k_c}{k} - s} dd \right) (1-d^2)(s^2-d^2)^{-2}(s^2-1) ~.
\end{align}

\begin{figure}
    \centering
    \includegraphics[width=0.4\linewidth]{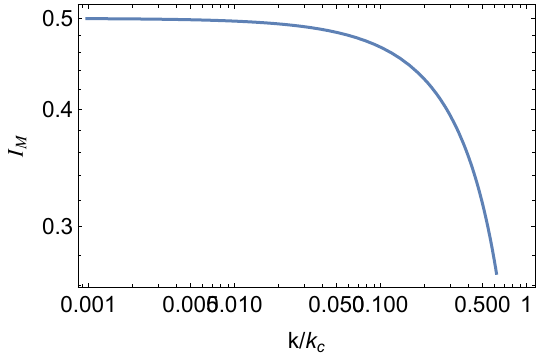}
    \includegraphics[width=0.4\linewidth]{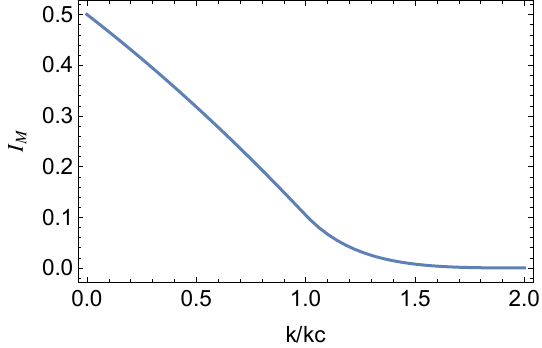}
    \caption{The momentum integral as a function of $k/k_c$. \zm{We demonstrate $\mathcal{I}_M$ for the range $k/k_c < 1$ in the left panel and $k/k_c \leq 2$ in the right panel. It is evident that $\mathcal{I}_M$ remains roughly a constant when $k/k_c \leq 0.1$ and quickly shrinks to 0 when $k/k_c > 0.1$.}}
    \label{fig:IM}
\end{figure}
\begin{figure}
    \centering
    \includegraphics[width=0.5\linewidth]{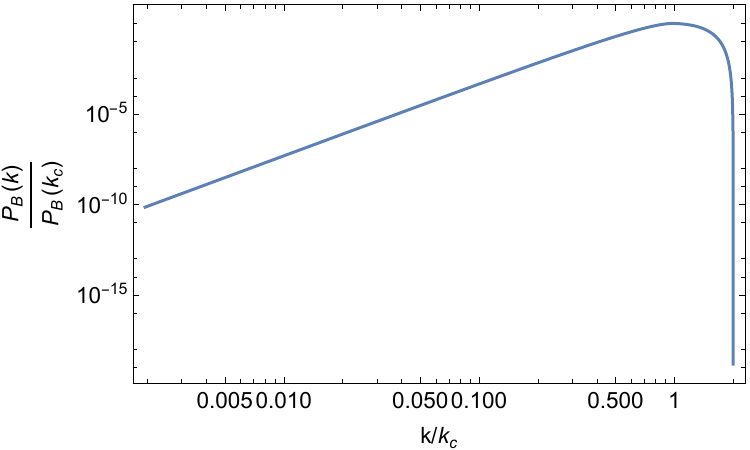}
    \caption{The magnetic power spectrum rescaled by $P_B(k=k_c)$.}
    \label{fig:PB}
\end{figure}

We numerically evaluate $\mathcal{I}_M$ and depict the result in Fig. \ref{fig:IM}. It is evident that $\mathcal{I}_M \simeq 1/2$ when $k/k_c \leq 0.1$, consistent with the analyze in Sec. \ref{sec:toy}, and decreases to 0 when $k/k_c$ grows. In the case $\mathcal{I}_M \simeq 1/2$, the magnetic power spectrum is
\begin{equation}
    P_B \simeq 5.0 \times 10^{-11} \varrho^2 \left( \frac{k}{a} \right)^4 \frac{v^4}{\lambda^2 m^2 T_2^2} ~,
\end{equation}
which is a highly blue spectrum with spectra index 4. When $k/k_c \geq 1$ it's difficult to compute the magnetic power spectrum analytically. We numerically evaluate $P_B$ and show the result in Fig. \ref{fig:PB}. We find that the magnetic power spectrum has an approximate spectra index 4 when $k < k_c$, and almost vanishes on the scale $k > k_c$. Therefore, the full magnetic power spectrum can be approximately written as
\begin{equation}
    P_B \simeq 5.0 \times 10^{-11} \varrho^2 \left( \frac{k}{a} \right)^4 \frac{v^4}{\lambda^2 m^2 T_2^2} \Theta (k_c - k) ~.
\end{equation}

\zm{It would be convenient to write the resultant magnetic power spectrum in terms of $\rho_{\phi}$, the energy density of scalar field, to discuss the back-reaction of both scalar field and magnetic fields. First, from the action \eqref{eq:action} and the effective potential \eqref{eq:Veffinteraction}, we see the scalar fields acquires an effective mass
\begin{equation}
    m_{\rm eff}^2 = \lambda^2 M_p \chi + \frac{\pi^2}{6} \varrho^2 T^2 + \varrho^2 A_{\mu}A^{\mu} ~.
\end{equation}
The term $\varrho^2 A_{\mu}A^{\mu}$ represents the back-reaction of the induced magnetic fields to the scalar field $\phi$, which is second-order and should be negligibly small. This can be verified by a simple estimation. The effective mass from the $A_{\mu}A^{\mu} |\phi|^2$ term is proportional to the energy density of induced magnetic fields $\rho_B$, which is of order $\mathcal{O}(10^{-72} [{\rm GeV}]^4)$ to be compatible with blazar observation as we will show in Sec. \ref{sec:electricfield}. On the other hand the temperature in radiation dominated epoch is $T > T_{\rm eq} = 10^{-9} [{\rm GeV}]$, thus the effective mass term $\varrho^2 A_{\mu}A^{\mu}$ is much smaller than $\frac{\pi^2}{12} \varrho^2 T^2$. In addition, after the tachyonic amplification, the auxiliary field $\chi$ rolls into the local vacuum $\langle \chi \rangle = M^2/\lambda^2 M_p$, so the effective mass after tachyonic amplifications is
\begin{equation}
    m_{\rm eff}^2 \simeq M^2 + \frac{\pi^2}{6} \varrho^2 T^2 ~,
\end{equation}
and the energy density of scalar field is accordingly
\begin{equation}
    \frac{d\ln \rho_{\phi}}{d\ln k} =  P_{\phi} \times  \begin{cases}
        \frac{\pi^2}{12} \varrho^2 T^2 & \varrho T  \gg  M \\
        \frac{\lambda^2 M^2}{2} & \varrho T \ll  M
    \end{cases}  ~.
\end{equation}
When $\varrho T \gg M$, $\rho_{\phi}$ scales as $a^{-4}$, and when $\varrho T \ll M$, $\rho_{\phi}$ scales as $a^{-3}$, similarly to a cold dark matter (CDM). We define 
\begin{equation}
    r \equiv \frac{\rho_{\phi}}{\rho_{\rm CDM}} ~,
\end{equation}
to be the ratio of $\rho_{\phi}$ versus $\rho_{\rm CDM}$, the energy density of CDM. Now, $r$ either decreases when $\rho T \gg M$ or remains constant when $\varrho T \ll M$, so $r$ reaches its maximum at $T = T_2$, the end of tachyonic phase. In addition, $\rho_{\rm CDM} \leq \rho_{\rm bg}$ in RD epoch. We conclude that $r(T=T_2) \ll 1$ is a sufficient condition for the energy density of $\phi$ to be negligibly small than the background one. This condition translates to
}
\begin{align}
    1 \gg r(T_2) & \nonumber \simeq \frac{P_\phi N}{3 H_{\rm eq}^2 M_p^2 (T_2/T_{\rm eq})^3} \times 
    \begin{cases}
         \frac{\pi^2}{6} \varrho^2 T_2^2 & \varrho T  \gg  M \\
         M^2  & \varrho T \ll  M
    \end{cases} \\
    & = \frac{5 \varrho^2 v^2N}{2T_2 T_{eq}} \times 
    \begin{cases}
         1 & \varrho T  \gg  M \\
        \frac{6 M^2}{\pi^2 \varrho^2 T_2^2}  & \varrho T \ll  M
    \end{cases} ~.
\end{align}
where $N$ is the e-folding number of inflation resulting from the momentum integration $d\ln k$ with respect to a scale-invariant power spectrum. \textcolor{blue}{In the following, we will use the shorthand notation $r \equiv r(T_2) $ as an abuse of notation. Note, that while the amplitude of the scalar field gets highly enhanced during the tachyonic phase, the scalar field energy density can still be highly subdominant $r \ll 1$ as long as the mass is sufficiently low $M \lesssim  \varrho T$ as the energy density scales as $r \propto \varrho^2$. In particular, the scaling is the same as for the magnetic power spectrum $P_B \propto \varrho^2$. Indeed, }
in terms of $r$, the magnetic power spectrum is
\begin{align}
\label{eq:PBbefore}
    P_B = 2.0 \times 10^{-11} \left( \frac{k}{a} \right)^4 \frac{v^2}{ m^2} \frac{r}{N} \frac{T_{eq}}{T_2} \Theta (k_c - k) \times 
    \begin{cases}
         1~,  & \varrho T  \gg  M \\
        \frac{\pi^2 \varrho^2 T_2^2}{6M^2} ~,  & \varrho T \ll  M
    \end{cases} ~.
\end{align}

\subsubsection{Coherence length and connection to observations}
\label{sec:coherence}
The key parameter in magneto-genesis is the coherence length, estimated by the wavelengths of magnetic fields averaged by the energy density:
\begin{equation}
    \lambda_B \simeq \frac{\int \frac{d^3k}{(2\pi)^3} \lambda_k \rho(k)}{\int \frac{d^3k}{(2\pi)^3} \rho(k)}  = \frac{\int \frac{d^3k}{(2\pi)^3} \frac{2\pi a}{k} P_B(k)}{\int \frac{d^3k}{(2\pi)^3} P_B(k)}  ~.
\end{equation}
Since the magnetic field scales as $P_{B} \propto (k/a)^4 \Theta (k_c - k)$, the coherence length is
\begin{equation}
\label{eq:lambdaB}
    \lambda_B  \simeq \frac{\int_0^{k_c} 2\pi k^2dk \left( \frac{k}{a} \right)^{3} }{\int_0^{k_c} k^2dk \left( \frac{k}{a} \right)^{4}} = \frac{7}{6}  \frac{2\pi a}{k_c} ~. 
\end{equation}

The blazar observations suggest a preferred range of the strength of magnetic fields with a coherence length of a few Mpc \cite{Neronov:2010gir,Finke:2015ona,MAGIC:2022piy}:
\begin{equation}
\label{eq:observation}
    B_{\rm today} \sim 10^{-15} - 10^{-17} G ~,~ \lambda_{B, \rm today} \sim 1 {\rm Mpc} ~.
\end{equation}
Thus, to be consistent with observations, we have
\begin{equation}
    \frac{7}{6}  \frac{2\pi a_{\rm today}}{k_c} \sim 1 {\rm Mpc} ~,~ \frac{k_c}{a_{\rm today}} \simeq 1.4 \times 10^{-1} {\rm Mpc}^{-1} ~.
\end{equation}
In light of the expression of $k_c = 5.9 \times 10^{25} \varrho {\rm Mpc}^{-1}$, implying a suggested model parameter
\begin{equation}
    \varrho \simeq 2.3 \times 10^{-27} ~.
\end{equation}
The integrated magnetic field strength is estimated by the root mean square of magnetic power spectrum:
\begin{equation}
    B_{\rm RMS}^2 = \int \frac{dk}{k} P_B(k) ~.
\end{equation}
From the expression of magnetic power spectrum \eqref{eq:PBbefore}, it's evident the magnetic power spectrum takes its maximum when $\varrho T_2 \gg M$, namely the effective thermal mass being much larger than the normal mass. For convenience let's work in this case, where the induced magnetic field is:
\begin{align}
    B^2 & \nonumber = \int \frac{dk}{k} 2.0 \times 10^{-11} \left( \frac{k}{a} \right)^4 \frac{v^2}{ m^2} \frac{r}{N} \frac{T_{eq}}{T_2} \Theta (k_c - k) \\
    & \nonumber = \int_0^{k_c} \frac{dk}{k} 2.0 \times 10^{-11} \left( \frac{k}{a} \right)^4 \frac{v^2}{ m^2} \frac{r}{N} \frac{T_{eq}}{T_2} \\
    & = 5.0 \times 10^{-12} \left( \frac{k_c}{a} \right)^4 \frac{v^2}{ m^2} \frac{r}{N} \frac{T_{eq}}{T_2} ~.
\end{align}
Therefore, the integrated magnetic field today is
\begin{equation}
    B({\rm today}) \leq 9.2 \times 10^{-65} \left( \frac{k_c/a_{\rm today}}{0.14 {\rm Mpc}^{-1}} \right)^2 \left( \frac{r/N}{10^{-2}} \right)^{\frac{1}{2}} \frac{v}{ m} \left( \frac{T_2}{T_{eq}} \right)^{-\frac{1}{2}} [{\rm G}] ~.
\end{equation}
We adopt an extreme parameter set $v = M_p$, $T_1 = 10T_2$ and $H_{\rm inf} = 10^{12} {\rm GeV}$, which surely makes an overestimation of $B$, then \eqref{eq:vlambdam} tells $c_3/T_2 = 4.5$, which in turn gets the value of $ m$. The resultant magnetic field is
\begin{equation}
    B({\rm today}) < 2.9 \times 10^{-32} \left( \frac{k/a_{\rm today}}{1 {\rm Mpc}^{-1}} \right)^2 \left( \frac{r/N}{10^{-2}} \right)^{\frac{1}{2}} \left(  \frac{ T_2 }{{\rm [GeV]}} \right)^{-\frac{5}{2}} \left( \frac{g_{\ast}}{106.75} \right)^{-\frac{1}{2}} [{\rm G}] ~,
\end{equation}
which has an upper bound $2.9 \times 10^{-22} [{\rm G}]$ when we take $T_2 = T_a = 10^{-4} {\rm [GeV]}$, marginally possible for seed magnetic fields. Notice that the $g_{\ast}$ dependence emerges from the definition of $c_3$ \eqref{eq:c3}. We conclude that the magnetic field generated during the tachyonic phase is too small to account for observations even after our overestimation.

\subsection{Magnetic field generated after the tachyonic phase}
There will be magneto-genesis after the tachyonic phase as well. To estimate the scalar power spectrum after $T = T_2$, we will simply assume that the transition of \zm{field} $\chi$ to the new vacuum is instantaneous. Therefore, the scalar power spectrum after the transition is estimated as $P_{\phi}(T = T_2^+) \simeq v^2$. The transfer function becomes
\begin{equation}
    \mathcal{T}(T;T_2) = \frac{\phi_k(T)}{\phi_k(T_2)} = \left( \frac{T}{T_2}\right)^n ~,~ T_3 < T < T_2 ~,
\end{equation}
where $T_3$ is the termination of magneto-genesis and $n = 1$ for $\varrho T \gg \lambda M$ case while $n = 3/2$ for $\varrho T \ll \lambda M$ case. Here we ignore the oscillations and only focus on the decaying part of transfer function. The time integral becomes
\begin{equation}
    \mathcal{I}(T_3,T_2;T_2) = \int^{T_3}_{T_2} -\frac{3\sqrt{10}} {2\pi \gamma \sqrt{g_{\ast}}} M_p (aT)_{\text{RD}} \frac{dT}{T^{4-2n} T_2^{2n}} ~.
\end{equation}
Assuming $T_3 \ll T_2$, we get
\begin{equation}
    \vert \mathcal{I}(T_3,T_2;T_2) \vert \simeq  \frac{3\sqrt{10}} {2\pi \sqrt{g_{\ast}}} \frac{M_p (aT)_{\text{RD}} }{\gamma T_2^{3} } \times
    \begin{cases}
         \frac{T_2}{T_3}  & n=1 \\
         \log \frac{T_2}{T_3} & n=3/2
    \end{cases}
    ~,
\end{equation}
and accordingly
\begin{equation}
    P_{B} = \frac{45 \varrho^2}{g_{\ast} \pi^2 } \left( \frac{k}{a} \right)^4 \frac{(aT)_{\text{RD}}^2 v^4 M_p^2}{\gamma^2 T_2^6 } \times
    \begin{cases}
        \frac{T_2^2}{T_3^2} & n=1 \\
        \log^2 \frac{T_2}{T_3} & n=3/2
    \end{cases}
    ~.
\end{equation}

In terms of $r$, the magnetic power spectrum becomes 
\begin{equation}
    P_{B} = \frac{18}{\pi^2 g_{\ast}} \left( \frac{k}{a} \right)^4 \frac{(aT)_{\text{RD}}^2 v^2 M_p^2 T_{eq}}{\gamma^2 T_2^3 T_3^2} \left( \frac{r}{N} \right) \times
    \begin{cases}
        1 & n=1 \\
        \frac{\pi^2 \varrho^2 T_2^2}{6\lambda^2M^2} \frac{\log^2 \frac{T_2}{T_3}}{T_2^2/T_3^2} & n=3/2
    \end{cases}
    ~.
\end{equation}
In the case $n = 3/2$, we have $\varrho^2 T_2^2 \ll  M^2$. Additionally, $\log (T_2/T_3)/(T_2/T_3) < 1$ since $T_2 > T_3$. We conclude that the induced magnetic field also reaches its maximum when $n = 1$, namely $\varrho T > \lambda M$. So we will focus on the case $n=1$, where the induced magnetic field is
\begin{align}
    & \quad \nonumber B = \sqrt{\frac{9}{2\pi^2 g_{\ast}} \left( \frac{k_c}{a} \right)^4 \frac{(aT)_{\text{RD}}^2 v^2 M_p^2 T_{eq}}{\gamma^2 T_2^3 T_3^2} \left( \frac{r}{N} \right)} \\
    & = 1.6 \times 10^{-23} \left( \frac{k_c/a_{\rm today}}{0.14{\rm Mpc}^{-1}} \right)^2 \left( \frac{r/N}{10^{-2}} \right)^{\frac{1}{2}} \left( \frac{v}{M_p} \right) \left( \frac{T_2}{1 {\rm MeV}} \right)^{-\frac{3}{2}} \left( \frac{T_3}{0.1 {\rm MeV}} \right)^{-1} [{\rm G}] ~,
\end{align}
marginally acceptable even if we've adopted the extreme parameter set $v = M_p$ and $T_3 = T_a$. We conclude that our formalism cannot generate enough cosmic magnetic field before the $e^+e^-$ annihilation epoch. Nonetheless, it's inspiring to see that we're quite close to sizable seed magnetic fields. Naturally, we would expect that after the annihilation epoch, the electric conductivity drops about ten orders of magnitude, and accordingly the induced magnetic field could be ten orders larger. Intuitively, sufficient magnetic field can be generated if the magnetogenesis happens after the annihilation epoch. We will verify this assertion in the next section.

\section{Magnetogenesis after the annihilation epoch}
\label{sec:MFafter}

\subsection{Magnetic field generated during the tachyonic phase}
The magnetic field generated after the annihilation epoch can be evaluated in a same way. We also start with the magnetogenesis during the tachyonic phase, where
\begin{align}
    \mathcal{I}(T_2,T_1;T_2) \equiv - \int^{T_2}_{T_1} \frac{(aT)_{\rm RD}^2}{2\kappa^2} \frac{dT}{T^3}  \mathcal{T}_{k_a}(T;T_2) \mathcal{T}_{k_b}(T;T_2) ~. 
\end{align}
Implemented with the transfer function \eqref{eq:transfermg}, we have
\begin{align}
    & \quad \mathcal{I}(T_2,T_1;T_2) \nonumber = \int^{T_2}_{T_1} -5.0 \times 10^{29} \left( \frac{g_{\ast}}{106.75} \right)^{-\frac{1}{2}} \frac{dT}{T_2^3} e^{\frac{2c_3^2}{T^2}} e^{-\frac{2c_3^2}{T_2^2}}  \\
    & \simeq  \frac{1.2 \times 10^{29}}{c_3^2} \left( \frac{g_{\ast}}{106.75} \right)^{-\frac{1}{2}} = \frac{7.3 \times 10^{9}}{ m [{\rm GeV}]} ~.
\end{align}
The power spectrum of magnetic field is 
\begin{align}
\label{eq:Bwithoutr}
    P_B &  = 1.1 \times 10^{20} \varrho^2 \frac{v^4}{m^2 {\rm [GeV]^2}} \left( \frac{k}{a} \right)^4 ~.
\end{align}
Given a typical coherence length $\lambda_B \simeq 1 {\rm Mpc}$, we have $k_c/a_{\rm today} = 0.14 {\rm Mpc}^{-1}$ and $\varrho = 2.4 \times 10^{-27}$, the integrated magnetic field is
\begin{equation}
\label{eq:Btachyonafter}
    B({\rm today}) = 5.1 \times 10^{-76} \left( \frac{\varrho}{2.4 \times 10^{-27}} \right) \left( \frac{k_c/a_{\rm today}}{0.14{\rm Mpc}^{-1}} \right)^2 \frac{v^2}{m[{\rm GeV}]} [{\rm G}] ~.
\end{equation}
We shall also check the back-reaction effect of magnetic field. The induced magnetic field gives negligible back-reactions if $r/N \ll 1$. After a quick examination we find that the magnetic field today also reaches its maximum when $\varrho T_2 \gg \lambda M$, where
\begin{equation}
    r/N = \frac{5\varrho^2 v^2}{2T_2T_{\rm eq}} ~.
\end{equation}
The value of $m$ has to be determined through \eqref{eq:vlambdam}. Since we assume the tachyonic phase happens after the annihilation epoch, we must have $T_1 \leq 0.1 {\rm MeV}$. We set $H_{\rm inf} = 10^{12} {\rm GeV}$ and $T_1 = 0.1 {\rm MeV}$ and organize the corresponding magnetic field from different model parameters in Table. \ref{tab:Bduring}. Note that the $g_{\ast}$ dependence is implicitly in the ratio $v/m$ according to \eqref{eq:vlambdam}. We see that a sizable PMF can be generated during the tachyonic phase with different model parameters. \zm{Specifically, seeds magnetic fields with $B \geq 10^{-21} [{\rm G}]$ can be produced for $v \geq 10^{12} {\rm GeV}$ and appropriate $T_2$ during the tachyonic phase}. In addition, it is suggested that the value of $v$ shall exceed $10^{14} {\rm GeV}$ to generate a PMF of $\mathcal{O}(10^{-16}) [{\rm G}]$ size during the tachyonic phase.
\begin{table}[ht]
\resizebox{\textwidth}{!}{%
\begin{tabular}{|c|c|c|c|c|c|c|}
\hline
\multicolumn{3}{|c|}{Model parameters}                  & \multicolumn{2}{|c|}{Auxiliary variables} & \multicolumn{2}{|c|}{Observables} \\ \hline
$ m /[{\rm GeV}]$ & $T_2 /[{\rm GeV}]$ & $\varrho$ & $v /[{\rm GeV}]$   &     $r/N$    &  B/[{\rm G}] & $\lambda_B$/[{\rm Mpc}] \\ \hline

$1.7 \times 10^{-29}$  & $10^{-5}$  &  $2.4 \times 10^{-27}$  &  $10^{12}$    & $1.4 \times 10^{-15}$   & $3.0 \times 10^{-23}$ & 1.0 \\ \hline

$1.7 \times 10^{-31}$  & $10^{-6}$  &  $2.4 \times 10^{-27}$  &  $10^{12}$    & $1.4 \times 10^{-14}$   & $3.0 \times 10^{-21}$ & 1.0  \\ \hline

$3.4 \times 10^{-29}$  & $10^{-5}$  &  $2.4 \times 10^{-27}$  &  $10^{13}$    & $1.4 \times 10^{-13}$   & $1.5 \times 10^{-21}$ & 1.0  \\ \hline

$3.4 \times 10^{-31}$  & $10^{-6}$  &  $2.4 \times 10^{-27}$  &   $10^{13}$    & $1.4 \times 10^{-12}$   & $1.5 \times 10^{-19}$ & 1.0  \\ \hline

$4.9 \times 10^{-29}$  & $10^{-5}$  &  $2.4 \times 10^{-27}$  &    $10^{14}$    & $1.4 \times 10^{-11}$   & $1.0 \times 10^{-19}$ & 1.0  \\ \hline

$4.9 \times 10^{-31}$  & $10^{-6}$  &  $2.4 \times 10^{-27}$  &  $10^{14}$    & $1.4 \times 10^{-10}$   & $1.0 \times 10^{-17}$ & 1.0  \\ \hline
\end{tabular}
}
\caption{Numerical result for our magneto-genesis scenario with different value of $v$ and $T_2$. Here, we set $H_{\rm inf} = 10^{12} {\rm GeV}$, $g_{\ast} = 106.75$ and $T_1 = 0.1 {\rm MeV}$.}
    \label{tab:Bduring}

\end{table}

Still, there are theoretical constraints that shall be considered. First, \eqref{eq:vlambdam} indicates a lower bound of $v$ once $H_{\rm inf}$, $T_1$ and $T_2$ are given. For a typical value $H_{\rm inf} = 10^{12} {\rm GeV}$, $T_1 = 0.1 {\rm MeV}$ and $T_2 = 0.01 {\rm MeV}$, $v$ has a lower bound of order $10^{10} {\rm GeV}$. Additionally, the tachyonic growth can happen only if $c_1^2/c_2 \ll 1$. Using \eqref{eq:c1c2condition}, this condition translates into $4.2 \times 10^{19} \varrho^2 [{\rm GeV}]/m \ll 1$ for large scale fluctuations. Since we fixed $\varrho = 2.4 \times 10^{-27}$ to ensure $\lambda_B = 1 [{\rm Mpc}]$, this condition further leads to the constraints $m \gg 2.4 \times 10^{-34} [{\rm GeV}]$. We confirm that all parameters in Table. \ref{tab:Bduring} meet this criteria. Furthermore, we find that $r/N \ll 10^{-10}$ due to the smallness of $\varrho$. We conclude that the energy density of scalar field is much smaller than that of CDM in our scenario, so the scalar field $\phi$ shall neither introduces severe back-reaction issues nor change the predictions of standard $\Lambda$CDM paradigm.

\subsection{Magnetic field generated after the tachyonic phase}
\label{sec:MFafteree}
After the tachyonic phase, the time integral is
\begin{align}
    & \nonumber \quad \mathcal{I}(T_3,T_2;T_2) = \int^{T_3}_{T_2} -\frac{(aT)_{\rm RD}^2}{2\kappa^2} \frac{dT}{T^{3-2n}T_2^{2n}} \\
    & = \frac{(aT)_{\rm RD}^2}{2\kappa^2} \times 
    \begin{cases}
         T_2^{-2} \ln \frac{T_2}{T_3} & \varrho T  \gg  M \\
        T_2^{-3} (T_2 - T_3)  & \varrho T \ll  M
    \end{cases} ~.
\end{align}
For $T_3 \ll T_2$ the result is not very sensitive on the precise value of $T_3$ and both cases are of the same order. For simplicity, we neglect the relative difference as it will be smaller than the errors resulting from our assumptions leading to
\begin{align}
\label{eq:PBafter}
    P_B \nonumber & \simeq 5.0 \times 10^{59} \left( \frac{g_{\ast}}{106.75} \right)^{-1} \varrho^2 \left( \frac{k}{a} \right)^4 v^4 T_2^{-4} \nonumber \\
    & \simeq 1.8 \times 10^{-151} \left( \frac{g_{\ast}}{106.75} \right)^{-1} \left( \frac{\varrho}{2.4 \times 10^{-27}} \right)^2 \left( \frac{k/a_{\rm today}}{0.14{\rm Mpc}^{-1}} \right)^4 \frac{v^4}{T_2^4}  {[\rm GeV]^4} ~,
    %\label{eq:Magnetic_Power_spectrum}
\end{align}
The integrated magnetic field is 
\begin{equation}
\label{eq:Bmaximum}
    B({\rm today}) = 3.4 \times 10^{-56} \left( \frac{g_{\ast}}{106.75} \right)^{-\frac{1}{2}} \left( \frac{\varrho}{2.4 \times 10^{-27}} \right) \left( \frac{k_c/a_{\rm today}}{0.14{\rm Mpc}^{-1}} \right)^2 \frac{v^2}{T_2^2} [{\rm G}] ~.
\end{equation}

Once we fixed $\rho = 2.4 \times 10^{-27}$ to get $\lambda_B = 1 {\rm Mpc}$, the magnetic field is solely determined by the ratio $v/T_2$. To get an intergalactic magnetic field with strength $B = \mathcal{O}(10^{-16}) [{\rm G}]$, it is suggested that $v/T_2 \simeq 10^{20}$. We organize the magnetic field from \eqref{eq:Bmaximum} in Table. \ref{tab:Bafter}. Notably, for the parameter set $v = 10^{14} {\rm MeV}$, $T_2 = 10^{-6} {\rm MeV}$, the induced magnetic fields after the tachyonic phase is about 30 times larger than that during the tachyonic phase. Thus, we may safely ignore the contributions during the tachyonic phase and conclude that sizable PMF of $\mathcal{O}(10^{-16}) [{\rm G}]$ can be produced after the tachyonic phase with reasonable parameter sets. It should be emphasized that we have set $g_{\ast} = 106.75$ in the numerical evaluation, however, this value may alter in different cosmic epochs and that would bring us a factor of order $\mathcal{O}(1)\sim\mathcal{O}(10)$ error, which does not invalidate our magneto-genesis scenario. 
\begin{table}[ht]\resizebox{\textwidth}{!}{%
\begin{tabular}{|c|c|c|c|c|c|c|}
\hline
\multicolumn{3}{|c|}{Model parameters}                  & \multicolumn{2}{|c|}{Auxiliary variables} & \multicolumn{2}{|c|}{Observables} \\ \hline
$ m /[{\rm GeV}]$ & $T_2 /[{\rm GeV}]$ & $\varrho$ & $v /[{\rm GeV}]$   &     $r/N$    &  B/[{\rm G}] & $\lambda_B$/[{\rm Mpc}] \\ \hline

$6.3 \times 10^{-29}$  & $10^{-5}$  &  $2.4 \times 10^{-27}$  &  $10^{15}$    & $1.4 \times 10^{-9}$   & $3.4 \times 10^{-16}$ & 1.0  \\ \hline

$4.9 \times 10^{-31}$  & $10^{-6}$  &  $2.4 \times 10^{-27}$  &  $10^{14}$    & $1.4 \times 10^{-10}$   & $3.4 \times 10^{-16}$ & 1.0  \\ \hline

$3.4 \times 10^{-33}$  & $10^{-7}$  &  $2.4 \times 10^{-27}$  &  $10^{13}$    & $1.4 \times 10^{-11}$   & $3.4 \times 10^{-16}$ & 1.0  \\ \hline

\end{tabular}
}
\caption{Numerical result for our magneto-genesis scenario with different value of $v$ and $T_2$. Here, we set $H_{\rm inf} = 10^{12} {\rm GeV}$, $g_{\ast} = 106.75$ and $T_1 = 0.1 {\rm MeV}$.}
\label{tab:Bafter}
\end{table}

\subsection{Discussions on conceptual issues}
\label{sec:electricfield}
In this section, we discuss several potential conceptual problems in our magneto-genesis scenario. 

First, we come to the potential over-production of electric fields. One may worry about the overproduction of electric field due to the drop of electric conductivity. This can be argued as follows. In plasma physics, one adopts the Debye length to characterize the electrostatic effect. We may assume that on scales much larger than Debye length, the universe is in a quasineutral state and the electric field is negligible. The lower limit of Debye length is estimated as 
\begin{equation}
    \lambda_D(T) \geq \sqrt{\frac{\epsilon_0 k_B T}{n_e q_e^2}} ~,
\end{equation}
where ion terms are dropped for simplicity. After annihilation the number density of electrons are
\begin{equation}
    n_e \simeq 10^{-10} n_{\gamma} = 10^{-10} \times \int \frac{8\pi}{(2\pi)^3} \frac{ E^2dE}{e^{E/T} - 1} = 2.4 \times 10^{-11} T^3 ~.
\end{equation}
We remind the readers that we're working in Planck units such that $\hbar = k_B = c = 1$. This leads to 
\begin{equation}
    \lambda_D = \frac{6.7 \times 10^5}{T} = 8.3 \times 10^4 \left( \frac{T}{0.1 {\rm MeV}} \right)^{-1} {\rm m} ~, 
\end{equation}
which is much smaller than galaxy scale. Thus, we are justified to work in this phase without the problem of the overproduction of electric field. 

Then, we discuss whether the total magnetic energy density remains subdominant to the background energy density. Since in our scenario $P_B$ is highly blue on large scales, and it peaks at $k \simeq k_c$, we may estimate the magnetic energy density by collection the contributions with $k \leq k_c$:
\begin{equation}
    \rho_B \simeq \int_0^{k_c} d\ln k \rho_k = \int_0^{k_c} d\ln k P_B(k) = \int_0^{k_c} \frac{dk}{k} \left( \frac{k}{k_c} \right)^4 P_B(k_c) = \frac{P_B(k_c)}{4} ~.
\end{equation}
Using \eqref{eq:PBafter}, along with the numerical results in Table \ref{tab:Bafter}, we see that when the integrated magnetic field $B \simeq 3.4 \times 10^{-16} [{\rm G}]$, the model parameters satisfy $v/T_2 = 10^{20}$, and $P_B(k_c) = 1.8 \times 10^{-71} {[\rm GeV]^4}$. As a result, the energy density of magnetic field $\rho_B$ is of $\mathcal{O}(10^{-72}) {[\rm GeV ]^4}$. On the other hand, the background energy density at matter-radiation equilibrium epoch is $\rho_{\rm eq} \simeq 9.7 \times 10^{-38} {[\rm GeV ]^4}$. Thus, $\rho_B \ll \rho_{\rm eq}$ and we verify that the total magnetic energy density is subdominant to the background one.

\zm{Utilizing the above result, we can also compute the ratio $\rho_B / \rho_{\phi}$. Notably, the energy density of magnetic field decays no slower than the scalar-field energy density, thus the ratio $\rho_B/\rho_{\phi}$ reaches its maximum at the end of tachyonic phase, $T = T_2$, and we compute 
\begin{equation}
    \frac{\rho_B(T_2)}{\rho_{\phi}(T_2)}  = \frac{\rho_B(T_2)}{\rho_{\gamma}(T_2)} \frac{\rho_{\gamma}(T_2)}{\rho_{\phi}(T_2)}  \sim 10^{-19} r^{-1} ~,
\end{equation}
which is quite small. For instance, in the second row of Table 2, $T = 10^{-6} {\rm GeV}$ and $r = 1.4 \times 10^{-10}$, thus in this case $\frac{\rho_B}{\rho_{\phi}} (T = T_2) \simeq 10^{-9}$. We conclude that the energy density of induced magnetic fields is way smaller than the energy density of scalar fields.}

Last, let us shortly comment about the non-linear evolution of the magnetic field which has been extensively discussed in \cite{PhysRevD.70.123003,Subramanian_2016,PhysRevLett.114.075001}. During radiation domination the magnetic field can get significantly damped at the peak frequency while the coherence length growths. However, the non-linear interaction only becomes important if the Alfven crossing time of a mode is smaller than the Hubble horizon, i.e.
\begin{align}
    k V_A(k) \tau \geq 1
\end{align}
where
\begin{align}
    V_A(k) = \sqrt{\frac{{\rm d} \rho_B}{{\rm d} \log k} \frac{1}{\rho_\gamma + p_\gamma} }
\end{align}
Using \eqref{eq:PBafter} the modes at the peak frequency $k_c$ are still far outside the non-linear regime during radiation matter equality. After recombination the non-linear interaction only provides logarithmic correction which can be neglected for our purposes \cite{PhysRevD.70.123003}. Note that our case differs from standard magnetogenesis where the non-linear effects are important (see for instance \cite{PhysRevD.96.083511}) as the modes are enhanced on deep superhorizon scales.

\section{Conclusion}
\label{sec:conclusion}
We propose a novel mechanism of primordial magneto-genesis that takes place in the standard radiation dominated epoch after the electroweak symmetry breaking phase to evade the baryon isocurvature  
problem and the over-production of electric fields. The conformal invariance of the EM field is broken because of its effective mass term that originated from the interaction with a scalar field. The scalar field is assumed to be very light and is extremely weakly coupled, and thus the strong coupling problem is absent. We find that sufficient magnetic fields capable to account for seed magnetic fields can be produced without the back-reaction problem provided that the scalar field experiences a tachyonic growth phase. Therefore, our mechanism provides a viable mechanism for generating large scale magnetic field without all above problems. 

This work highlights the possibility of primordial magneto-genesis in radiation dominated epoch and motivates further exploration along this line. Notably, in our minimal setup, the spectra index of the magnetic field power spectrum $P_B(k)$ is always 4. As future astrophysical observations may reveal more information about the spectra index, it would be interesting to generalize our models and explore the allowed spectra index in future works.

\acknowledgments
We would like to thank G. Domènech, M. Sasaki for their helpful comments and discussions.
C.L. is supported by the grant No. 2021/42/E/ST9/00260 from the National Science Centre, Poland. M. Z. is supported by the Fundamental Research Funds for the Central Universities Grant No. YJ202551, and the Grant No. 2021/42/E/ST9/00260 from the National Science Centre, Poland. A. G. receives support by the DFG under the Emmy-Noether program, project number 496592360, and by the JSPS KAKENHI grant No. JP24K00624.

\bibliography{mf.bib}
\bibliographystyle{JHEP}

\end{document}